\documentclass[11pt,twocolumn,twoside]{article}
\usepackage{style}

\setlist{nolistsep}
\crefalias{subappendix}{section}
\definecolor{Gray}{gray}{0.9}

\graphicspath{{img/}}
\begin{document}
\onecolumn
\title{\DS{}: A lung cancer screening dataset from Québec for reliable AI evaluation}
\author[1,2,3,4, *]{Johanna Topalis} 
\author[6, *]{Yannick Lemaréchal}
\author[6]{Gabriel Couture}
\author[6]{Hanife Ersay}
\author[1]{Loran Nas}
\author[1,3]{Balthasar Schachtner}
\author[1,4,7]{Bastian Oliver Sabel}
\author[1,2,3]{Michael Ingrisch}
\author[5,6,+]{Philippe Després}
\author[1,3,4]{Katharina Jeblick}

\affil[1]{Department of Radiology, LMU University Hospital, LMU Medizin, Ludwig-Maximilians-Universität München, Munich, Germany}
\affil[2]{relAI – Konrad Zuse School of Excellence in Reliable AI, Munich, Germany}
\affil[3]{Munich Center for Machine Learning (MCML), Munich, Germany}
\affil[4]{Comprehensive Pneumology Center (CPC-M), Member of the German Center for Lung Research (DZL), Munich, Germany}
\affil[5]{Département de physique, de génie physique et d’optique, Université Laval, Québec, Canada}
\affil[6]{Centre de recherche de l'Institut universitaire de cardiologie et de pneumologie de Québec\,-\,Université Laval, Québec, Canada}
\affil[7]{Department of Radiology, Asklepios Lung Clinic Munich-Gauting, Germany}
\affil[*]{Shared first authorship}
\affil[+]{Corresponding author}

\maketitle

\vspace{1.5cm} 
\begin{tcolorbox}[
 colback=gray!10,
  colframe=gray!10,
  boxrule=0.5pt,
  arc=0mm
]
\begin{abstract}
\absdiv{BACKGROUND}
AI-based tools are expected to play a key role in lung cancer screening (\LCS). 
The evaluation of these tools often relied on private datasets or fragmented public datasets with inconsistent availability of clinical data and  expert nodule annotations. 
This has hindered reproducibility, comparability, and independent assessment of AI tools. 
To address these gaps, we present \DS{}, a novel, large, comprehensively annotated dataset derived from LDCT lung cancer screening in Québec, which is accessible for reliable AI evaluation. \\
\absdiv{METHODS}
\DS{} was curated and is hosted on PARADIM, a digital infrastructure supporting DICOM data de-identification, annotation, long-term storage and analysis. 
\LDCT{} scans and clinical information were combined to a curated collection and complemented by voxel-level, expert-derived pulmonary nodule and cyst segmentations, and Lung-RADS-based lesion characteristics.
Key dataset characteristics and statistics were reported, including participant demographics, imaging parameters, radiological findings, and segmented lesion characteristics.  Dataset analyses were conducted according to PARADIM's model-to-data approach, in which user code is  executed on the platform such that data remains on premise. \\
\absdiv{RESULTS}
\DS{} currently comprises 1,678 \LDCT{} baseline studies from 1,678 participants. Most studies were classified as Lung-RADS 2. A total of 1,776 pulmonary nodules and 60 cysts were annotated, with a median (interquartile range) volume of $89\,(56,\,159)\,\mathrm{mm}^3$ and $1,384\,(589,\,3793)\,\mathrm{mm}^3$ respectively. \\
\absdiv{CONCLUSIONS}
Compared to publicly available \LCS{} datasets,  \DS{} currently represents the largest accessible annotated real-world \LCS{} dataset.  Access to \DS{} is provided via a model-to-data approach, enabling privacy-preserving, reproducible, and independent real-world evaluation of AI tools in \LCS. \\
\end{abstract}
\end{tcolorbox}

\twocolumn
\begin{refsection}
\section{Introduction}
Lung cancer is the leading cause of cancer-related mortality worldwide~\cite{bray2024global}. 
Early detection through low-dose computed tomography (LDCT) screening has been shown to reduce lung cancer mortality~\cite{de2020reduced, national2011reduced}.
Consequently, lung cancer screening (\LCS{}) programs have been implemented in several countries~\cite{interactive_map_lung_cancer_screening}. 
Over the past decade, the development of AI models for pulmonary nodule detection, segmentation, tracking, and lung cancer risk prediction
has seen significant progress in research~\cite{cellina2023artificial, de2023artificial, setio2017validation, peeters2026benchmarking, mikhael2023sybil} and industry~\cite{antonissen2025artificial}.
Given the potential of AI models to support clinicians at \LDCT{} scan interpretation, AI-based tools are expected to become a key component of \LCS{} programs~\cite{LuKrFrühErkV}.

However, the focus of \LCS{} AI research has mainly been the training of advanced models; 
their independent, comparable, and reproducible evaluation has received less attention~\cite{antonissen2025artificial, hahn2025requirements, khunte2023trends} and has suffered from a fragmented landscape of \LCS{} datasets. 
So far, evaluation has relied on either private datasets or on a limited number of publicly available (or controlled-access) datasets~\cite{nguyen2025scoping}, such as the LIDC-IDRI~\cite{armato2011lung}, LUNA16~\cite{setio2017validation}, and data from the U.S. national \LCS{} trial (NLST)~\cite{national2011reduced}.
Private datasets are typically inaccessible to external researchers, making it difficult to reproduce  results or compare the performance of different AI models~\cite{nguyen2025scoping}.
Public or controlled-access datasets, on the other hand, are (easily) accessible, but they also come with limitations. 
For example, CT scans in LIDC-IDRI/LUNA16 were not exclusively derived from  \LCS{}~\cite{armato2011lung} and lack screening-relevant metadata (e.\,g., Lung-RADS~\cite{christensen2024acr} classification). 
In contrast, the standard NLST imaging release does not include expert lung nodule segmentations, limiting its utility for evaluating, e.\,g., AI-based nodule detection, segmentation, and tracking tools.
In addition, the integrity of evaluation results on public \LCS{} datasets is generally not guaranteed (e.\,g., test data might have been (inadvertently) used during training~\cite{van2024comparison, bassi2024touchstonebenchmarkrightway}).

This situation has resulted in several challenges for reliable evaluation of AI-based tools for \LCS:
(1) the fact that some \LCS{} datasets, including LIDC-IDRI/LUNA16, originate (in large parts) from clinical routine rather than real-world \LCS{}s,
(2) the lack of \LCS{} datasets combining both expert lung nodule annotations and screening-specific clinical data, 
(3) limited reproducibility and comparability when inaccessible private datasets are used for evaluation, and
(4) the absence of accessible yet independent~\cite{fdamedicaldevices} \LCS{} evaluation datasets.

To address these challenges, we present a novel multi-center LUng Cancer screening dataset from QUÉbec, \DS{}, specifically designed for independent, comparable, and reproducible real-world evaluation of AI-based tools for \LCS{}. 
\DS{} comprises 1,678 baseline LDCT studies from the Lung Cancer Screening Project in Québec, Canada~\cite{screeningreport, quebec_screening_project}, acquired at two different hospitals. 
It provides comprehensive screening-specific information (e.\,g., PLCOm2012~\cite{tammemagi2013selection}, Lung-RADS scores~\cite{christensen2024acr}), as well as rich demographic, clinical, and imaging-related data. 
The imaging studies have been further enriched with expert reference segmentation masks and additional Lung-RADS-relevant characteristics for lung nodules and cysts. 

\begin{figure*}[t]
  \centering
   \includegraphics[width=1\textwidth]{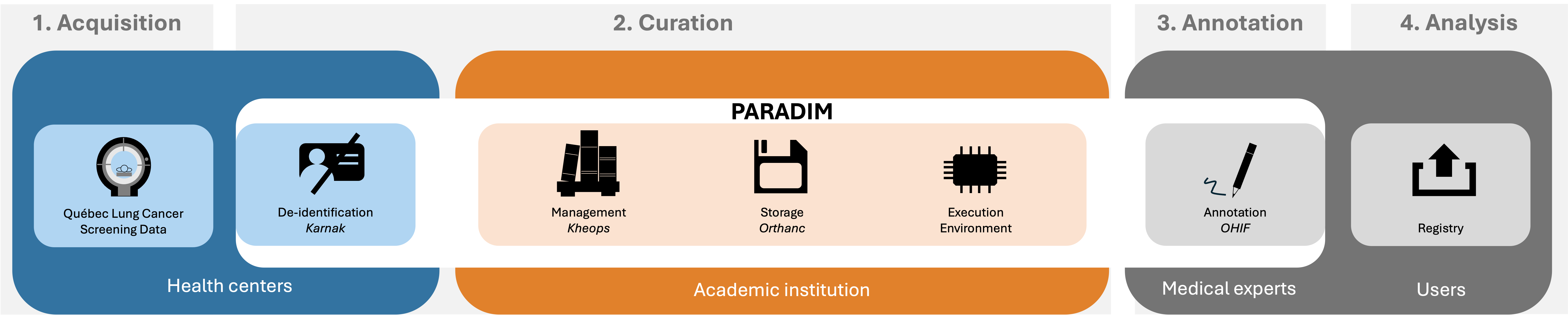}
  \caption{
  Workflow from data acquisition to analysis using PARADIM. PARADIM's components (white area) are deployed across different environments to facilitate seamless data processing. In the health centers (blue zone), lung cancer screening data is stored in local hospital information systems. Through PARADIM’s data pipelines, DICOM objects are automatically retrieved, de-identified (Karnak), and then transferred to the academic institution's network (orange zone), where they are stored in a DICOM server (Orthanc) and organized into collections (Kheops). DICOM images are annotated by medical experts using PARADIM's OHIF-based annotation pipeline~\cite{ohifdicomannotator}. 
  Users can upload containerized data analysis code to a registry and execute it on collections within PARADIM.  All data remains on premise; only aggregated analysis results can be extracted from the platform. Cf.~\cite{lemarechal2025paradim} for further details on PARADIM's architecture.} 
  \label{fig:paradim}
\end{figure*}

\DS{} was curated and hosted on PARADIM (Platform for the Annotation, Reuse, and Analysis of Medical Images)~\cite{lemarechal2025paradim}. 
PARADIM is a digital infrastructure supporting customizable DICOM de-identification, curation, annotation, organization and long-term storage. 
Additionally, it supports the automated execution of large-scale operations on image collections. 
Access to \DS{} on PARADIM follows a model-to-data approach: instead of granting external users direct access to the dataset, they upload their models, which are then exposed to the DICOM data on PARADIM within a managed execution environment. 
This design preserves patient privacy and ensures traceability and reproducibility of evaluation results, while also guaranteeing that no testing data is used for training. 
Provided formal approvals, \DS{} is accessible on PARADIM for evaluation purposes to both research and industry partners, providing a robust foundation for the reliable assessment of AI tools for \LCS{}.

\section{Materials and Methods}
The workflow  used to create and analyze \DS{} is summarized  in  Fig.~\ref{fig:paradim} and described in the following sections. 
It includes 1.~data acquisition, 2.~curation, 3.~annotation, and 4.~analysis. 
All steps were carried out with PARADIM's streamlined and largely automated pipelines. 
Data retrieval and analysis was approved by the ethics committee of {\em Centre de recherche de l'Institut universitaire de cardiologie et de pneumologie de Québec-Université Laval} (IUCPQ-ULaval), and required a Privacy Impact Assessment since informed consent was not possible for this project (Ethics approval 2023-3946, 22299).

\subsection{Data acquisition}
LDCT images and clinical information were derived from lung cancer screening in Québec, Canada (Supplemental Material~S1.1.1), which began in June 2021 as a demonstration project~\cite{screeningreport} and is now offered routinely~\cite{quebec_screening_project}. 
Eligible participants are current or former smokers (with cessation $\leq$ 15 years) aged 55-74 years with a $\geq20$ years smoking history and a PLCOm2012noRace score $\geq2\%$. Participants underwent standardized LDCT screening (Supplemental Material~S1.1.2) at designated centers, where radiologists evaluated the scans and assigned a Lung-RADS score (Supplemental Material~S1.1.3)  to guide follow-up imaging or further diagnostic examinations. 

LDCT scans and associated radiology reports were archived in local hospitals' Picture Archiving and Communication Systems (PACS). 
Participants' records were systematically collected and stored in a centralized REDCap~\cite{harris2009research}-based database. 
These included demographic information (e.\,g., sex, age, ethnicity), smoking history, medical history (e.\,g., comorbidities), the PLCOm2012noRace risk score at enrollment, and radiological findings (e.\,g., Lung-RADS score; Supplemental Material~S1.1.3) for each LDCT. 

\subsection{Data curation}
In an automated process, eligible LDCT scans were identified in the PACS of two participating hospitals, de-identified  using the DICOM gateway Karnak~\cite{osirixfoundationKarnak} and sent to PARADIM (Supplemental Material~S1.2.1). 
Participant's data from the centralized REDCap-based screening database was likewise de-identified, and a selection of information uploaded to PARADIM as DICOM Structured Reporting (SR) files (cf.  Tab.~S1 for included participants’ characteristics at enrollment and Tab.~S2 for included radiological findings). Further details on this data ingestion process are provided in Supplemental Material~S1.2.2.

All data was stored in DICOM format in PARADIM's central storage backend, Orthanc~\cite{jodogne2018orthanc}, an open-source DICOM server. 
In PARADIM, Orthanc is not directly exposed to users. 
Instead, all interactions with the stored data is mediated through Kheops~\cite{osirixfoundationKHEOPSManageShare}, a platform built for managing and sharing imaging data in research settings. 
It organizes the DICOM data into collections, i.e., logical sets  that aggregate DICOM series with tailored access permissions.
We created a Kheops collection for \DS{} (Fig.~\ref{fig:data_tree}), based on the selected  exclusion/inclusion principles (Supplemental Material~S1.2.3).

\subsection{Data annotation} 
\label{sec:meth:data_annotation}
All LDCT studies in \DS{} were annotated retrospectively by a radiologist with four years of experience in general radiology using PARADIM’s annotation pipeline~\cite{ohifdicomannotator}, which leverages the web imaging platform by the Open Health Imaging Foundation (OHIF)~\cite{ziegler2020open}. 
The annotations were performed in a blinded fashion using LDCTs only.
Information on the participants’ medical history and screening findings were not provided to the annotating radiologist. The latter was not provided to approximate a typical image-reading setting.
Before starting with the annotations, the radiologist was provided a detailed annotation guide (Supplemental Material~S4).

All detected pulmonary nodules (3\,mm $\leq$ diameter $\leq$ 30\,mm; consistent with~\cite{armato2011lung}) and tumor masses (diameter > 30\,mm)~\cite{austin1996glossary} as well as cysts were manually segmented at the voxel-level. 
Smaller nodules or cysts were also segmented if detected. 
Part-solid nodules were annotated in two parts, with the solid and non-solid components segmented separately and then linked as components of the same part-solid nodule.
In the following all segmented findings (pulmonary nodules and cysts) will be referred to as lesions. 
For each study, all lesions were stored in a single DICOM Segmentation (SEG) file. \\
The lesions were then further classified as solid nodule, non-solid nodule or cyst. 
Lesions were further characterized using binary labels (yes/no, no as initial selection) for key radiological features derived from Lung-RADS~v2022~\cite{christensen2024acr} (Supplemental Material~S1.3.1). 
The radiologist also identified the lung segment where the lesion was located. 
Nodule characteristics were saved as DICOM SR objects.
Annotation metadata, including annotator identity, annotation date, and software details, were embedded in DICOM tags of the DICOM SEG and SR files to ensure traceability, following FAIR~\cite{wilkinsonFAIRGuidingPrinciples2016} best practices for data documentation. \\
To assess the annotation quality, an intra- and inter-reader variability assessment was performed (Supplemental Material~S1.3.2).

\subsection{Data analysis}
We extracted key characteristics of \DS{} to describe the dataset in Section~\ref{sec:res}. 
Participants' characteristics, LDCT technical parameters (including scanner manufacturers, scanner model names, and reconstruction kernels) and radiological findings were analyzed using medians and interquartile ranges (IQRs) or by total counts. \\
For segmented lesions, we plotted the distribution of lesion counts per participant and reported the total count of solid, non-solid, part-solid nodules, and cysts, as well as the total count of each lesion characteristic.
In addition, we calculated the volumes ("Approximate volume" radiomic feature) of lesions directly from the baseline DICOM CT and corresponding SEG objects using PyRadiomics~\cite{van2017computational}. The code execution on the DICOM objects followed PARADIM's model-to-data approach described in more detail in Supplemental Material~S1.4.
We aggregated volume data over all studies to plot the lesion volume distribution, calculate the median (IQR), report the number of solid nodules exceeding volumetric size thresholds of $34\,\mathrm{mm}^3$, $113\,\mathrm{mm}^3$, $268\,\mathrm{mm}^3$ and $1,767\,\mathrm{mm}^3$ (as used in Lung-RADS v2022 for size-based risk stratification~\cite{christensen2024acr}), and the number of (solid) tumor masses with a volume $>14,137\,mm^3$ (corresponding to the volume of a sphere with a diameter $>3\,cm$).

\section{Results}
\label{sec:res}
\DS{}  comprises 1,678  baseline studies with 3,356 series from 1,678 participants, acquired between June 2021 and November 2025 at two hospitals in Québec, Canada. 
1,539 studies were acquired in center 1 and 139 at center 2.

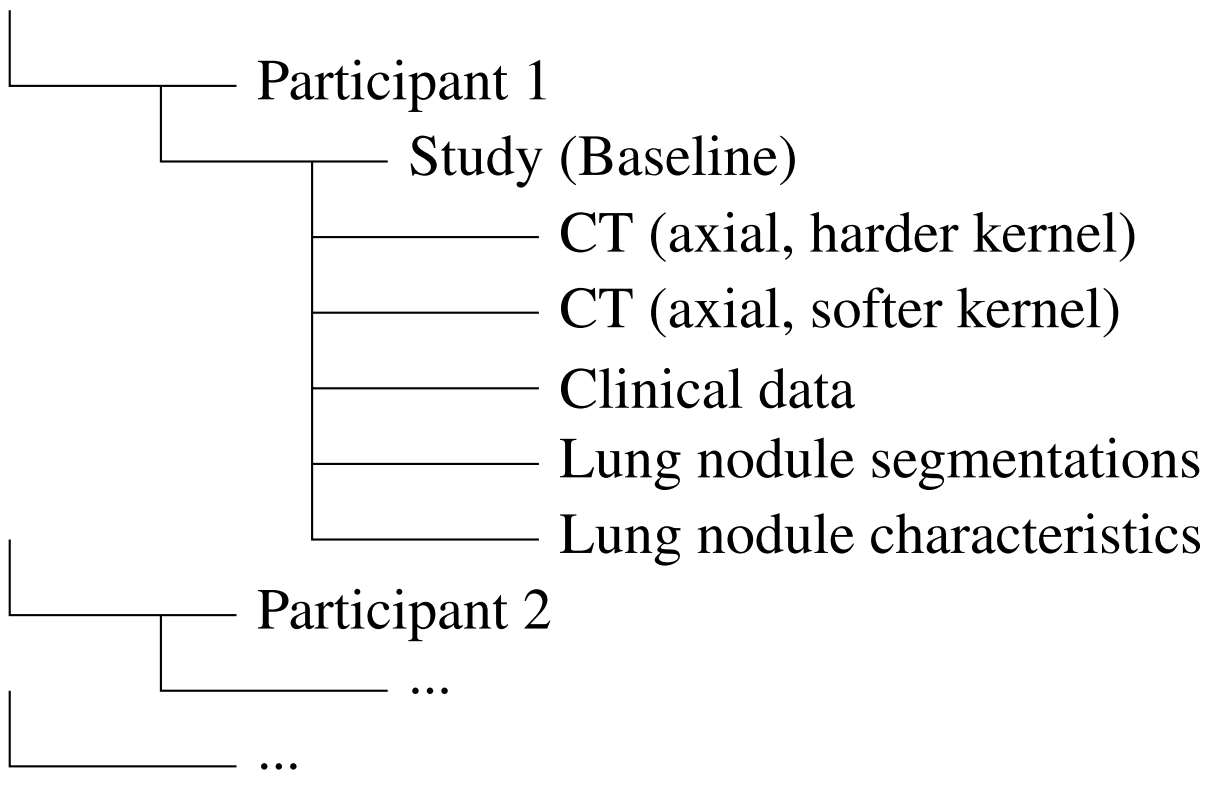
\begin{figure}[h!]
    \centering
\begin{tikzpicture}
    \treeroot{}
    \treeentry{Participant 1}{1}
    \treeentry{Study (Baseline)}{2}
        \treeentry{CT (axial, harder kernel)}{3}
        \treeentry{CT (axial, softer kernel)}{3}
        \treeentry{Clinical data}{3}
         \treeentry{Lung nodule segmentations}{3}
         \treeentry{Lung nodule characteristics}{3}
    \treeentry{Participant 2}{1}
    \treeentry{...}{2}
    \treeentry{...}{1}
\end{tikzpicture}
\caption{Overview of the data structure for \DS{}. \DS{} contains baseline  studies, each of which include two axial CTs (reconstructed with a softer and harder  kernel), and a Structured Report (SR) with clinical data (radiological findings and participants' characteristics). \DS{} is also enriched with lesion segmentations (saved as DICOM Segmentation (SEG)) and characteristics (saved as DICOM SR)).}
\label{fig:data_tree}
\end{figure}

Each study in \DS{} (Fig.~\ref{fig:data_tree}) includes two axial CT series reconstructed with a hard and a soft kernel, an SR with clinical data (participants’ characteristics and radiological findings from the clinical program database), a SEG with lesion segmentations and a corresponding SR with lesion characteristics.

\subsection{Participant’s characteristics}
\label{sec:res:part_char}
\begin{figure*}[t!]
\subfloat[]{%
    \includegraphics[width=.33\linewidth]{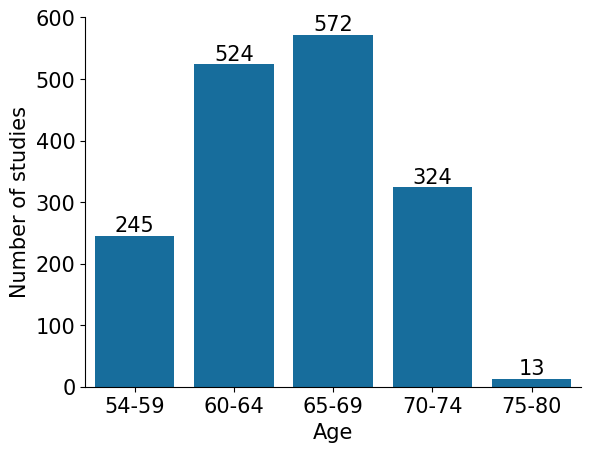}%
    \label{subfig:age}%
}\hfill
\subfloat[]{%
    \includegraphics[width=.33\linewidth]{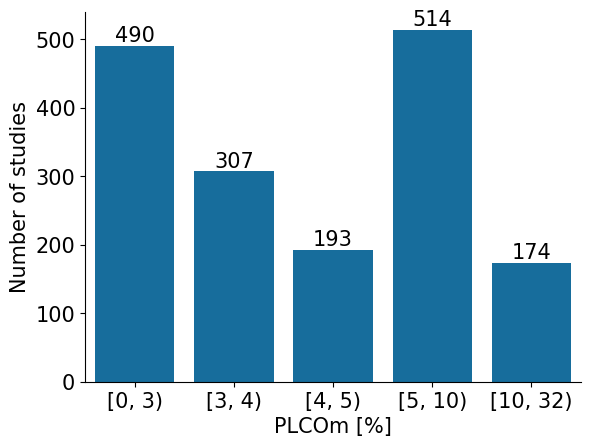}%
    \label{subfig:plcom}%
}\hfill 
\subfloat[]{%
    \includegraphics[width=.33\linewidth]{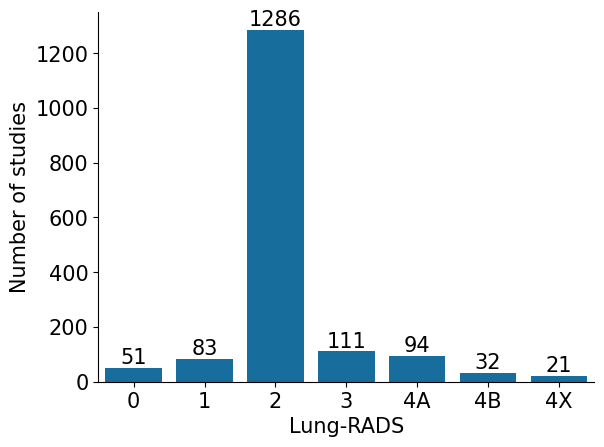}%
    \label{subfig:lungrads}%
} \hfill
\caption{Participants' (a) age (in years) at \LDCT{} imaging, (b) PLCOm2012noRace score, and (c) Lung-RADS.} 
\label{fig:cohort}
\end{figure*}

\DS{} contains 878 male and 800 female participants with a  median (IQR) age of $65\,(61,\,69)$ years (Fig.~\ref{subfig:age}). 
The majority of participants were Caucasian (1,649), but the dataset also includes individuals from other racial or ethnic backgrounds. 
1268 participants smoked, while 410 had quit smoking prior to enrollment. 
Chronic respiratory conditions were prevalent at enrollment, with 404 participants diagnosed with COPD, emphysema, or chronic bronchitis. 
The calculated median (IQR) PLCOm2012noRace score was 4.2~\% (2.7~\%, 7.0~\%) (Fig.~\ref{subfig:plcom}).
A summary of all baseline characteristics is provided in Tab.~S1.

\subsection{LDCT technical parameters}
\label{sec:results:ldct}
\LDCT{}  studies were acquired with Philips CT iCT (1,539) or Siemens SOMATOM Definition Edge (139).
Axial series were reconstructed with one softer and one harder kernel: L and YA for Philips, Br43f/Bf42f and Bl57f for Siemens, respectively.

\subsection{Radiological Findings}  
\label{sec:res:rad}  
Radiological findings available in \DS{} are summarized in Tab.~S2. Most studies were classified as Lung-RADS~2 (1,286), followed by Lung-RADS~3 (111). The highest Lung-RADS~4B and 4X were assigned for 32 and 21 studies, respectively (Fig.~\ref{subfig:lungrads}). 
In 1,637 participants, other radiological findings were reported. 
Emphysema were reported for 1,423 studies. 
In 1,490 studies, findings in the heart were reported, among these were 793, 376 and 214 studies with low, moderate and severe coronary calcification.

\subsection{Nodule characteristics}
\begin{figure*}[t!]
\centering
 \begin{center}

    \subfloat[]{
        \includegraphics[
            height=0.20\textwidth
        ]{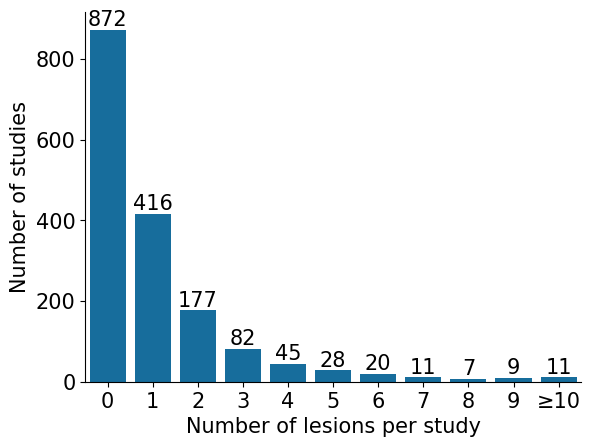}
        \label{fig:nodule_count_per_study}
    }
    \subfloat[]{
        \includegraphics[
            height=0.20\textwidth
        ]{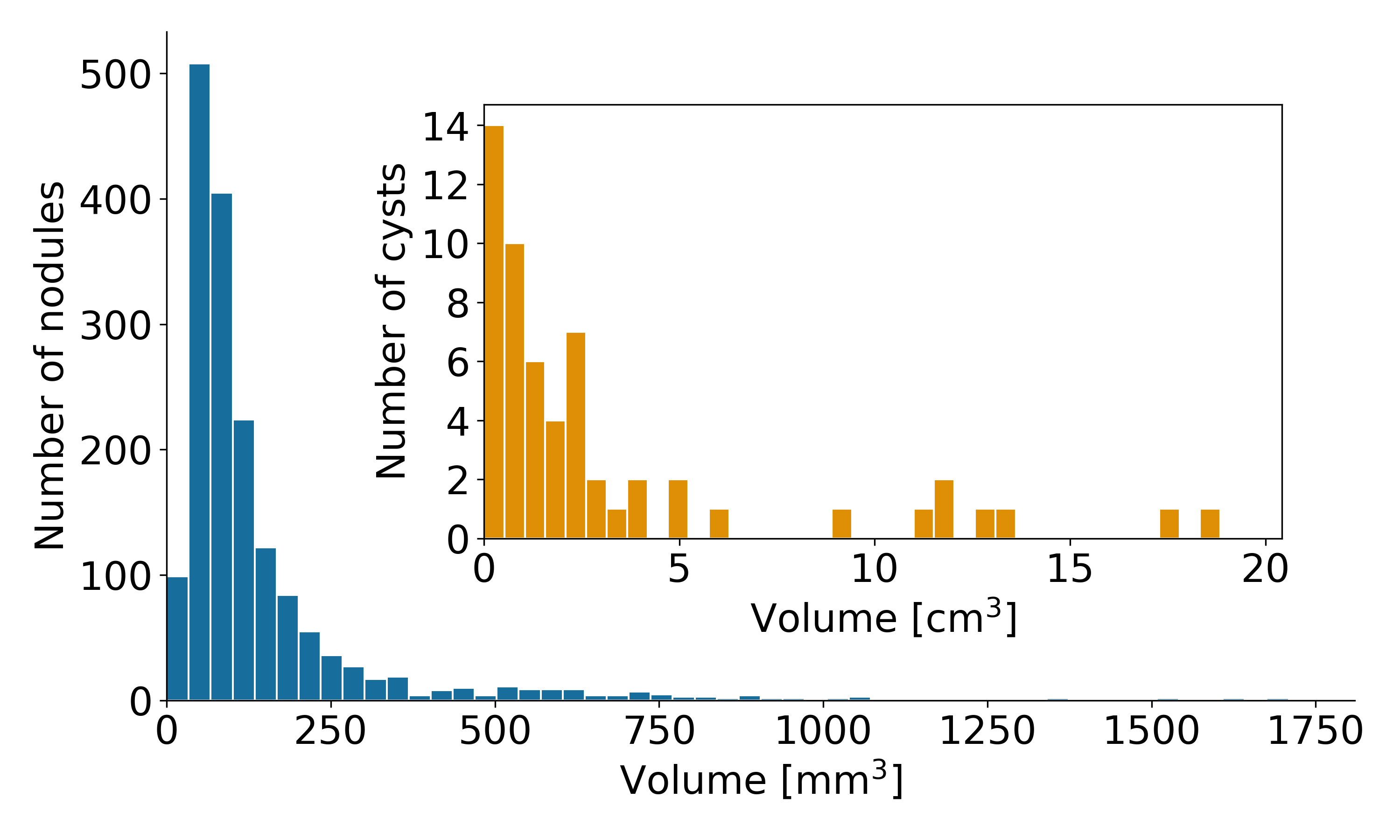}
        \label{fig:nodule_volume}
    }
    \\

\subfloat[]{
    \raisebox{7.5mm}{
        \includegraphics[width=0.22\textwidth]{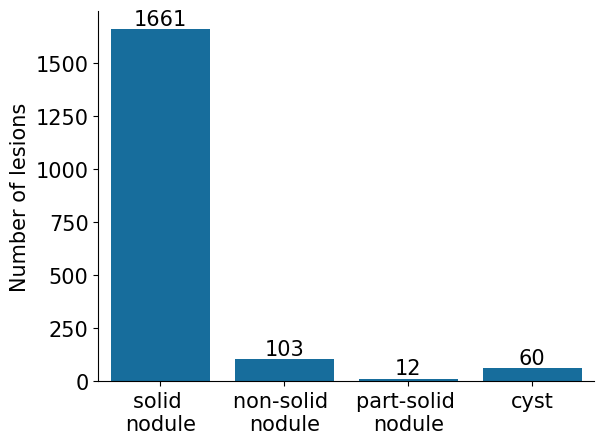}
    }
    \label{fig:nodule_type}
}
\hfill
\subfloat[]{
    \raisebox{0mm}{
        \includegraphics[width=0.35\textwidth]{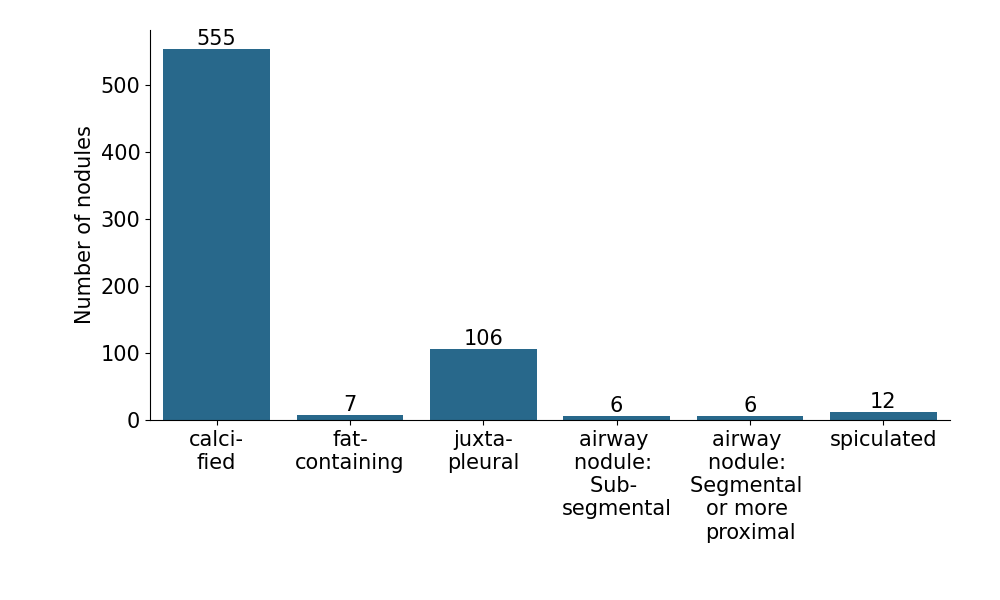}
    }
    \label{fig:solid_nodule_subtype}
}
\hfill
\subfloat[]{
    \raisebox{7.5mm}{
        \includegraphics[width=0.3\textwidth]{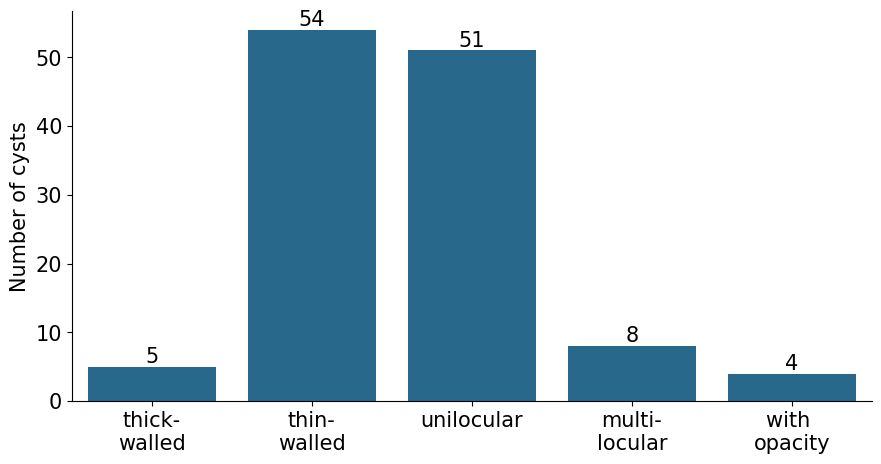}
    }
    \label{fig:atypical_cyst_subtype}
}
\end{center}
\caption{(a) Number of pulmonary nodules and cysts per study. 
(b) Volume distributions of nodules (blue; plotted until $1,800\,\mathrm{mm}^3$) and cysts  (orange; plotted until $20\,\mathrm{cm}^3$). 
(c) Number of findings classified as solid, non-solid or part-solid pulmonary nodules or cysts. 
(d) Number of  characteristics of solid pulmonary nodules. 
(e) Number of  characteristics of cysts. 
}
\label{fig:nodules_analysis}
\end{figure*}

In 872 baseline studies, the annotating radiologist did not detect any pulmonary nodules or cysts, whereas at least one lesion was annotated in 806 studies. 
The distribution of total lesions per study is illustrated in Fig.~\ref{fig:nodule_count_per_study}, with the number of annotated lesions ranging from 1 to 43 per study.
The majority of the detected 1,776 nodules were solid (1,661), followed by  non-solid (103) and part-solid (12) (Fig.~\ref{fig:nodule_type}). 
555 solid pulmonary nodules were labeled as complete, central, popcorn, or concentric ring calcifications, and 106 as juxtapleural nodules with smooth margins; and oval, lentiform, or triangular shape. Additionally, 12 nodules were annotated as as spiculated, 7 nodules as fat-containing, 6 as sub-segmental airway nodules and 6 as segmental or more proximal airway nodules (Fig.~\ref{fig:solid_nodule_subtype}). 
The volume distribution of all segmented nodules (and tumor masses) is shown in Fig.~\ref{fig:nodule_volume}. 
The median (interquartile range) nodule volume was $89\,(56,\,159)\,\mathrm{mm}^3$. 
The smallest detected nodule measured $17\,\mathrm{mm}^3$, while the largest mass had a volume of approximately $172\,\mathrm{cm}^3$. 
A total of 1,555 solid nodules exceeded the Lung-RADS-relevant volume threshold of $34\,\mathrm{mm}^3$, among which 571,  188 and 41 further had a volume greater than $113\,\mathrm{mm}^3$, $268\,\mathrm{mm}^3$ and $1,767\,\mathrm{mm}^3$, respectively. 
10 solid masses had a volume larger than $14,137\,\mathrm{mm}^3$.
In addition, 60 cysts were detected with 54 labeled as thin-walled and 5 as thick walled. 51 cysts were annotated as unilocular, 8 as multi-locular and 4 as cysts with opacity (Fig.~\ref{fig:atypical_cyst_subtype}). 
Example annotations illustrating pulmonary nodule and cysts characteristics are provided in Fig.~\ref{fig:nodules}. Intra- and inter-reader analysis results are presented in Supplemental Material~S2.1.

\newcommand{\lungfig}[1]{
  \subfloat[]{
    \includegraphics[
      width=.25\textwidth,
      height=.18\textwidth,
      keepaspectratio
    ]{#1}
  }
}
\begin{figure*}[t!]
\centering
\lungfig{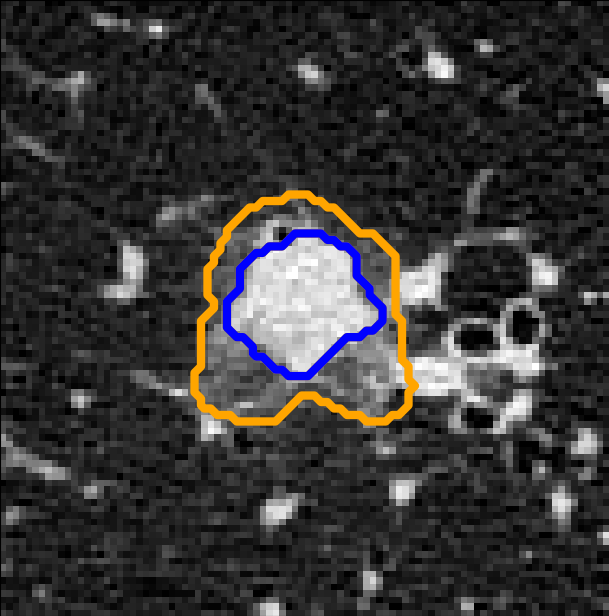}
\hspace{2pt}
\lungfig{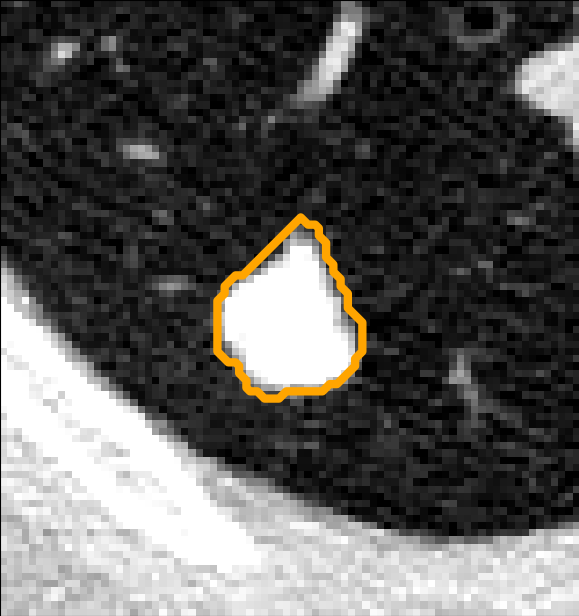}
\hspace{2pt}
\lungfig{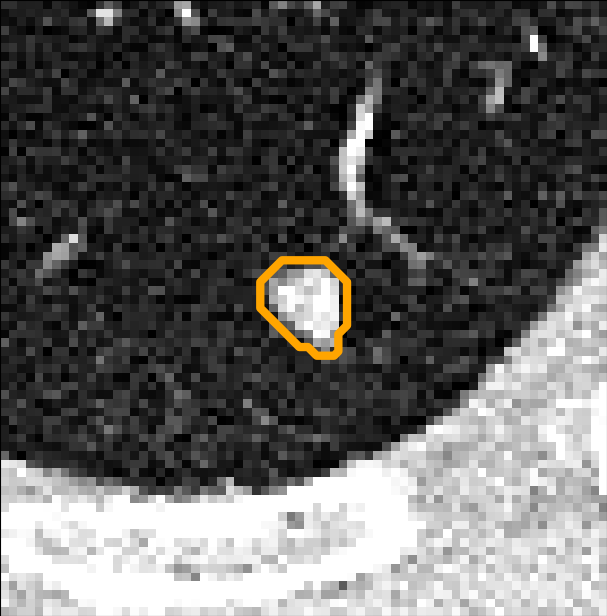}
\hspace{2pt}
\lungfig{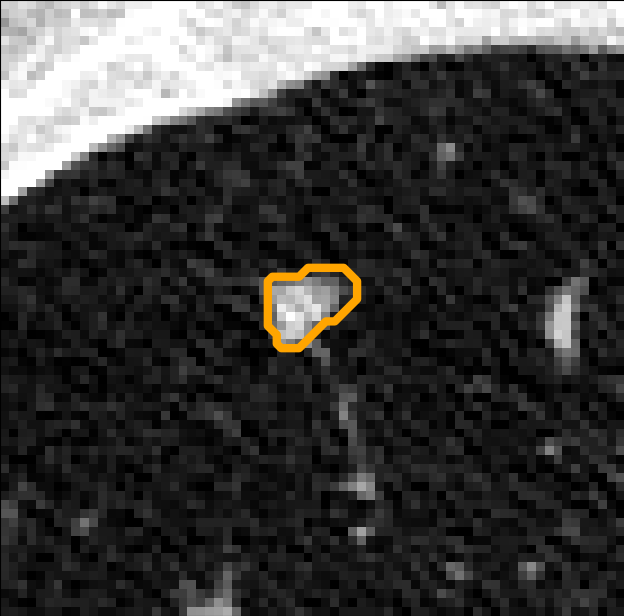}
\hfill\\

\lungfig{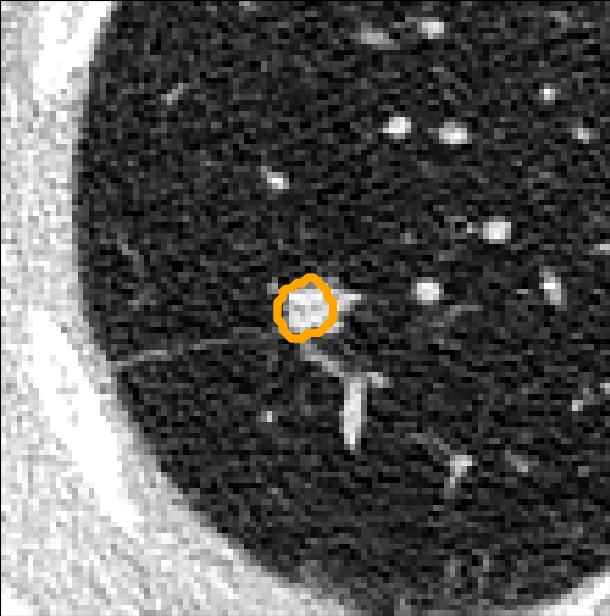}
\hspace{2pt}
\lungfig{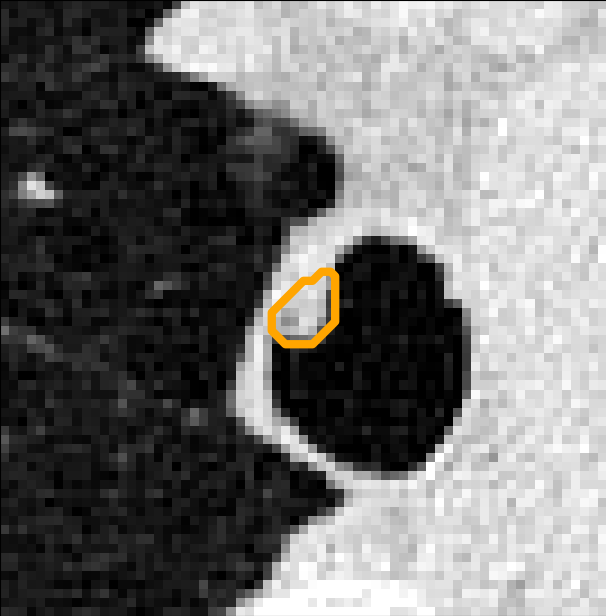}
\hspace{2pt}
\lungfig{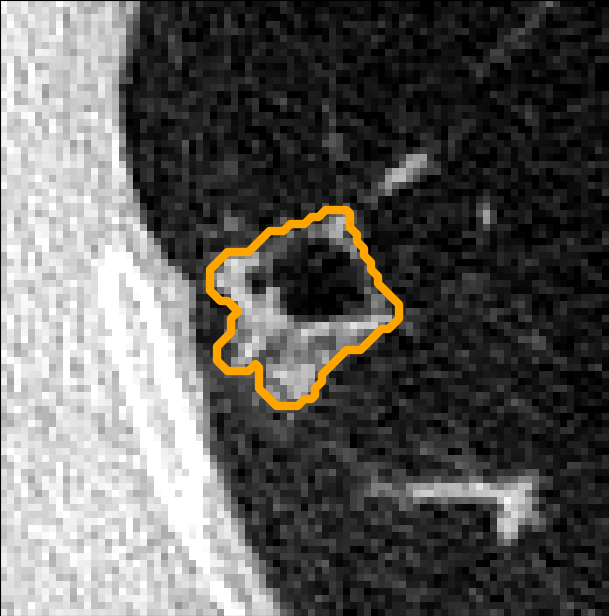}
\hspace{2pt}
\lungfig{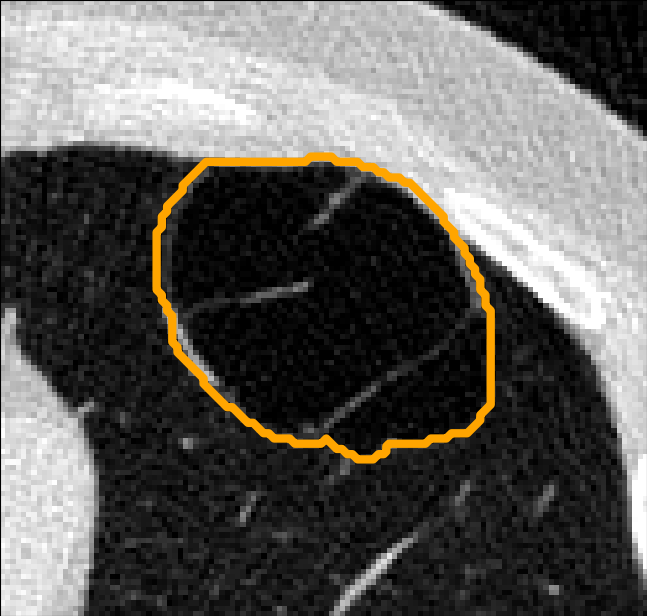}
\hfill\\

\lungfig{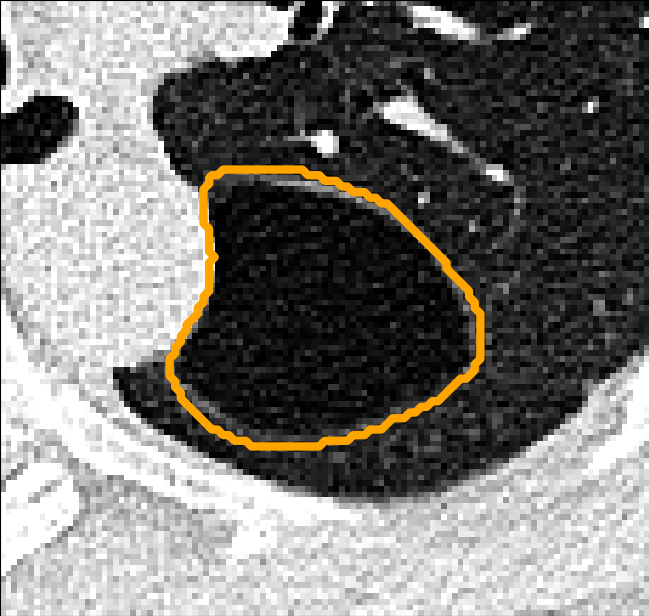}
\hspace{2pt}
\lungfig{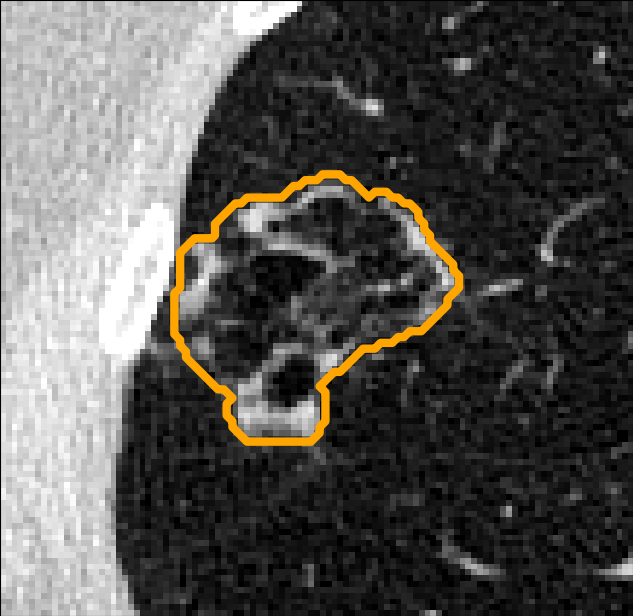}
\hspace{2pt}
\lungfig{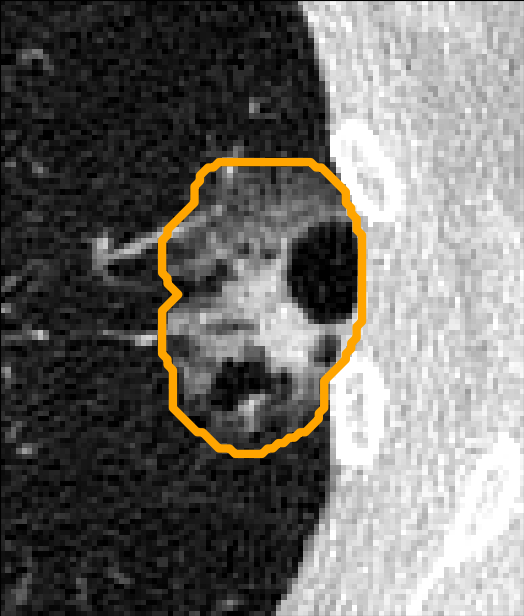}
\caption{
Exemplary annotations of the categories (a) part-solid nodule, (b) nodule with complete, central, popcorn, or concentric ring calcification, (c) fat-containing nodule, (d) juxtapleural nodule with smooth margins; and oval, lentiform, or triangular shape, (e) airway nodule: subsegmental, and (f) airway nodule: segmental or more proximal, (g) thick-walled cyst (>=2mm - symmetric, asymmetric, or nodular), (h) thin-walled cyst (uniform wall thickness < 2 mm), (i) unilocular cyst, (j) multilocular cyst and (k) cyst with opacity (nodular, ground glass, or consolidation).}
\label{fig:nodules}
\end{figure*}

\section{Discussion}
Given the increasing integration of AI support tools into  \LCS{} programs~\cite{LuKrFrühErkV}, reliable evaluation of these tools is critical for their safe and effective clinical adoption. 
As highlighted in the HANSE \LCS{} trial~\cite{kondrashova2024comparison}, differences between AI software tools can lead to clinically relevant discrepancies in key outputs such as nodule volume measurements and Lung-RADS classifications which may result in inconsistent participant management.
Yet, current performance data and fair comparisons, across both research and commercial models,  remain limited~\cite{hahn2025requirements, khunte2023trends, antonissen2025artificial}. 

In this study, we introduced a new \LCS{} dataset from Québec, \DS{}, designed to address this gap by enabling systematic evaluation of AI models across key \LCS{} tasks including nodule detection, segmentation/volumetric measurement, classification, and Lung-RADS scoring. 
\DS{} includes 1,678 baseline \LDCT{} studies, paired with rich screening-specific metadata, expert reference segmentations, and detailed Lung-RADS-relevant labels  for pulmonary nodules. 
These resources enable nuanced analyses across patient subgroups, and nodule characteristics (e.\,g., non-solid vs. solid), facilitating clinically relevant performance benchmarking. 

While several \LCS{} trials and programs have been implemented worldwide~\cite{interactive_map_lung_cancer_screening}, only a few datasets have been made accessible for AI development and evaluation. 
The most widely used datasets in the context of AI in \LCS{} include the LIDC-IDRI~\cite{armato2011lung} and its subset LUNA16~\cite{setio2017validation}, which were in large parts not acquired as part of a \LCS{}, but mainly in clinical routine. 
The NLST dataset~\cite{national2011reduced} from the large U.S. national trial was the first publicly released \LCS{} dataset, while the recently published Duke Lung Cancer Screening (DLCS) Dataset ~\cite{wang2025duke} is the latest addition, providing more contemporary data from the Duke University Health System. 

Compared to these datasets (Tab.~\ref{tab:dataset_comparison}), \DS{} includes the largest number of participants and studies after NLST, and is the largest \LCS{} dataset with expert-annotated nodules. 
In contrast to older datasets, \DS{} reflects recent imaging practices and CT technology. Similar to NLST and DLCS, it is drawn from a real-world \LCS{}, enhancing its relevance for evaluating AI tools in contemporary screening settings.

\DS{}  stands out for the richness of its nodule annotations, offering both voxel-level segmentations and detailed lesion classifications aligned with Lung-RADS v2022. While LIDC-IDRI includes segmentations from multiple annotators and LUNA16 provides derived nodule centers and diameters, DLCS relies on semi-automatically generated bounding boxes~\cite{wang2025duke}. 
In contrast, the standard NLST imaging release lacks exact nodule locations. Although LIDC-IDRI and NLST report some nodule characteristics, these are not aligned with Lung-RADS.
\DS{} and NLST include detailed participant characteristics relevant for screening, such as demographics, smoking, and medical history. In DLCS, only demographic data and smoking history are provided. 
Lung-RADS scores are provided only in \DS{} and DLCS.

In sum, to the best of our knowledge, \DS{} is the largest accessible \LCS{} dataset for AI evaluation that combines precise locations with Lung-RADS-based labels for pulmonary nodules and cysts, comprehensive participant characteristics relevant to \LCS{}, and detailed screening findings, including Lung-RADS scores and additional radiological findings. 

We host \DS{} on the PARADIM platform. Not only the \LDCT{} scans, but also nodule segmentations, lesion classifications, and clinical data are provided as  DICOM objects. 
The DICOM format enables the inclusion of all original imaging metadata (where not de-identified), as well as additional contextual information such as annotator identity, annotation time, and methods, ensuring clear data documentation. 
Crucially, DICOM tags (e.\,g., in SEG and SR objects) store references to the original CT series, enabling precise traceability of all data elements. 
While LIDC-IDRI and NLST also use DICOM for imaging data, LUNA16 and DLCS are only available in NIfTI format, limiting metadata depth and traceability.   

Unlike the other datasets in Tab.~\ref{tab:dataset_comparison}, which are fully downloadable (raising concerns about potential prior exposure to test data during pretraining or tuning~\cite{van2024comparison, bassi2024touchstonebenchmarkrightway}) \DS{} is only accessible through a model-to-data approach on the PARADIM platform. 
In this setup, users upload containerized AI models, which are then automatically executed within a secure compute environment against the DICOM data stored on PARADIM. 
These workflows are standardized, scalable, and reproducible, with all data, models, and outputs versioned, traceable and reusable.
This approach ensures that the underlying data remain unseen, preserving participant privacy while upholding the integrity of evaluation results. 
The platform thus enables fair and consistent comparisons across AI tools, directly addressing concerns around reproducibility, replicability, and independence in AI evaluation~\cite{bassi2024touchstonebenchmarkrightway, fdamedicaldevices}. 

While \DS{} fills a need in the AI evaluation landscape for \LCS, several limitations should be noted. 
First, \DS{} currently includes data from only two health centers in Québec, with not all participants from these centers yet included and no follow-up imaging, which may reduce representativeness of the full Québec screening cohort. 
Moreover,  annotations were provided by a single radiologist with only a subset being analyzed for intra- and inter-reader annotator variability (Supplemental Material~S3.2).
Finally, on the technical side, parts of the (non-imaging) data ingestion from the clinical program database into PARADIM still required some manual operations, affecting scalability. 

This work presents the first version of our dataset, i.e.,  \DS{}~v1.0.0.
To further enhance its value for AI evaluation,  \DS{}  will be continuously extended. New  dataset versions can be easily and efficiently created on PARADIM by adding imaging and clinical data or annotations using its streamlined and largely automated data ingestion, annotation, and curation pipelines. Upcoming versions of \DS{} will include additional participants from the ongoing Québec screening program, and complete follow-up studies for more comprehensive longitudinal coverage. Planned additions also include further clinical information such diagnostic results and annotations by multiple experts.
As a next step, we will evaluate AI models, e.\,g., for lung nodule detection, using LUCQUE.

To conclude, we presented \DS{}, a novel, large and  comprehensive and annotated multi-center LUng Cancer screening dataset from QUÉbec.
\DS{} was curated and is hosted on PARADIM, a digital infrastructure enabling streamlined and largely automated de-identification, curation, annotation, organization and long-term storage of medical data collections in DICOM format, as well as automated analysis of this data. Access to \DS{} on PARADIM for research and industrial purposes is possible through a model-to-data approach, enabling reliable real-world evaluation of AI-based tools for \LCS{}.

\begin{table*}[t!]
\centering
\caption{Comparison of \DS{} to other accessible datasets used in the context of \LCS{}. Further accessible datasets (cf. Supplemental Material~S3.1) are not listed in this table due to their low number of studies (< 300), missing \LCS{} context or detailed information. \\
$^{*1}$ Additional 448 scans were withheld as test cases and are not publicly available~\cite{wang2025duke}.
$^{*2}$ To the best of our knowledge, no acquisition dates for the LIDC-IDRI data are publicly available. As the paper~\cite{armato2011lung} was published in 2011, all images must have been acquired before 2011. 
$^{*3}$ Only partly \LDCT{}s from lung cancer screening~\cite{armato2011lung}. To the best of our knowledge, the exact number of screening and routine CTs in the dataset is not officially published, although the SeriesDescription field in the DICOM headers may provide some indication.  
$^{*4}$ For 8 patients, CTs at two different timepoints are available. 
$^{*5}$ Can be derived from the segmentations.
$^{*6}$ For non-calcified nodules or masses with $\leq$4~mm, NLST contains (if filled) the longest diameter and longest perpendicular diameter, the location (lung lobe) of the nodule epicenter, and the CT slice number containing the greatest diameter. 
$^{*7}$ Over $900$ NLST participants diagnosed with lung cancer within one year after imaging were (in the context of other studies) semi-automatically annotated with 2D~\cite{mikhael2023sybil} and 3D~\cite{tushar2024ai} bounding boxes. 
$^{*8}$ Assessment of subtlety, internal structure, calcification, sphericity, margin, lobulation, spiculation, texture, and malignancy of segmented lung nodules on a Likert scale from 1 to 5. 
$^{*9}$ For non-calcified nodules or masses with $\leq$4~mm diameter, NLST contains (if filled) a characterization of nodule margin and the predominant attenuation. 
$^{*10}$ Participants' characteristics includes comprehensive demographic information alongside detailed smoking and medical history relevant to \LCS{}. 
$^{*11}$ Nodule-based malignancy scoring from 1-5\,\cite{armato2011lung}.
}
\begin{tabular}{ l | l | l | l | l |  l }
&  \textbf{\DS{}} & \textbf{LIDC-IDRI}\,\cite{armato2011lung} &  \textbf{LUNA16}\,\cite{setio2017validation} &  \textbf{NLST}\,\cite{national2011reduced}&  \textbf{DLCS}\,\cite{wang2025duke} \\
\toprule
\rowcolor{Gray}
\textbf{Dataset composition} &  &  &  &  &  \\
No. patients/participants &  1,678 & 1,010 & 888 & 26,254 & 1,613$^{*1}$ \\
\rowcolor{Gray}
No. CT studies &  1,678 & 1,018 & 888 & 73,116 & 1,613$^{*1}$ \\
Image acquisition years &  2021-2025 & before 2011$^{*2}$ & before 2011$^{*2}$ & 2002-2004 & 2015-2021 \\
\rowcolor{Gray}
Real world \LCS{} CTs & \checkmark &  \crossmark$^{*3}$ &  \crossmark$^{*3}$ &  \checkmark & \checkmark \\
\makecell[l]{Studies at multiple \\ timepoints}  &  \crossmark & \crossmark$^{*4}$ & \crossmark$^{*4}$ & \checkmark & \crossmark \\
\hline
\rowcolor{Gray}
\textbf{Nodule annotations} & & & & & \\
Segmentations & \checkmark & \checkmark & \crossmark & \crossmark & \crossmark\\
\rowcolor{Gray}
Bounding boxes & \crossmark$^{*5}$  & \crossmark$^{*5}$  & \checkmark & \crossmark$^{*6,7}$& \checkmark \\
Nodule characteristics  &\checkmark & \checkmark$^{*8}$ & \crossmark & \checkmark$^{*9}$ & \crossmark \\
\rowcolor{Gray}
Annotation style & manually & manually & manually & \crossmark$^{*7}$ & semi-autom. \\
\makecell[l]{Multiple independent \\ annotators}  & \crossmark & \checkmark & \crossmark & \crossmark & \crossmark \\
\hline
\rowcolor{Gray}
\textbf{Participants' charac. $^{*10}$}  & \checkmark & \crossmark & \crossmark & \checkmark & \checkmark \\
\hline
\textbf{Radiological findings}  &  &   &  & & \\
\rowcolor{Gray}
Lung-RADS  & \checkmark & \crossmark$^{*11}$ & \crossmark& \crossmark & \checkmark\\
\makecell[l]{Additional radiological \\ findings} & \checkmark & \crossmark& \crossmark & \checkmark & \crossmark \\
\hline
\rowcolor{Gray}
\textbf{Data format}  & DICOM  & DICOM & NifTI & DICOM & NifTI \\
\hline
\textbf{Available for download}  & \crossmark & 
\checkmark 
& \checkmark 
& \checkmark 
& \checkmark$^{*1}$ 
\\
\bottomrule
\end{tabular}
\label{tab:dataset_comparison}
\end{table*}

\section*{Acknowledgements}
KJ was supported through funding from Bayerisches Staatsministerium für Wissenschaft und Kunst in cooperation with Fonds de recherche du Québec.
JT was supported by the DAAD programme Konrad Zuse Schools of Excellence in Artificial Intelligence, sponsored by the Federal Ministry of Research, Technology and Space.
This research was funded by Data Transformation Grants of the Canadian Cancer Society (CCS grants \#708520 and \#707586). Additional funding was provided by the Natural Sciences and Engineering Research Council of Canada (Discovery program grant RGPIN-2018-04588 and CREATE program grant 528124-2019), the Ministère de l’Économie, de l’Innovation et de l’Énergie du Québec and the Fonds de recherche du Québec (Collaborative Bilateral Research Program Bavaria-Québec, grant 322724). \\
Authors thank Pierre-Luc Asselin for his assistance in retrieving data from the clinical program database. \\
Generative AI was used as a supporting tool for copy-editing parts of the manuscript. The authors carefully reviewed and revised the content as necessary. The ideas expressed in this work are exclusively those of the human authors. The authors take full responsibility for the final manuscript.

\newpage
\section*{Authors contributions}
\textbf{JT:}  Conceptualization, Data Curation, Formal analysis, Project administration, Software, Validation, Visualization, Investigation,  Writing - Original Draft, Writing - Review \& Editing. 
\textbf{YL:} Data Curation, Software, Validation,  Writing - Review \& Editing. 
\textbf{GC:} Data Curation, Software, Validation, Writing - Review \& Editing. 
\textbf{HE:} Data Curation, Writing - Review \& Editing. 
\textbf{LN:} Data Curation, Writing - Review \& Editing. 
\textbf{BS:} Conceptualization, Methodology, Writing - Review \& Editing.
\textbf{BOS:} Validation, Writing - Review \& Editing.
\textbf{MI:} Funding acquisition, Project administration, Resources, Supervision,  Writing - Review \& Editing.
\textbf{PD:} Conceptualization, Funding acquisition, Project administration, Resources, Supervision, Writing - Review \& Editing. 
\textbf{KJ:} Conceptualization, Funding acquisition, Methodology, Project administration, Supervision, Validation, Writing - Original Draft, Writing - Review \& Editing.

\section*{Data Availability}
\DS{}~v1.0.0 is made accessible via PARADIM through research collaborations (corresponding author: PD).  Additional information and updates can be found on PARADIM's website \url{https://paradim.science/}.

\section*{Code Availability}
Code is made available on GitHub at \url{https://github.com/paradim-platform/lucque-lung-cancer-screening-dataset}.

\clearpage
\interlinepenalty=100000000
\clearpage
\raggedbottom
\printbibliography
\end{refsection}

\clearpage
\begin{refsection}
\setcounter{table}{0}
\renewcommand{\thetable}{S\arabic{table}}
\setcounter{figure}{0}
\renewcommand{\thefigure}{S\arabic{figure}}
\setcounter{section}{0}
\renewcommand{\thesection}{S\arabic{section}}

\section*{Supplemental Material}
\label{sec:appendix}
\definecolor{RoyalBlue}{rgb}{0.0, 0.14, 0.4}

\section{Methods}
\subsection{Data acquisition}
\label{sec:appendix:data_acquisition}
\subsubsection{Québec lung cancer screening project}
\label{sec:appendix:data_acquisition-project}
The Québec lung cancer screening demonstration trial aimed to screen 3,000 participants over two annual rounds~\cite{screeningreport2, martel2024ep}. 
The project has since transitioned into a provincial \LCS{} program~\cite{quebec_screening_project}. 
Data in \DS{} originated from both the demonstration project and the ongoing regional \LCS{} program.

\subsubsection{LDCT acquisition protocol}
\label{sec:appendix:data_acquisition-acquisition}
LDCT scans were acquired at participating hospitals following a standardized acquisition and reconstruction protocol designed for \LCS{}. Imaging was performed during breath-hold at the end of the inspiration phase. 
A CTDI\textsubscript{vol} dose below 2 mGy was defined for standard-weight participants to adhere to the low-dose requirement for screening.
The acquisition protocol was standardized across scanners, with parameters including rotation time, collimation, pitch, CT x-ray tube voltage, dose reduction index, tube current modulation, and CTDI\textsubscript{vol} tailored to the specific hardware in use. 
Images were reconstructed in axial, coronal, and sagittal planes.
Axial images were reconstructed with both a softer kernel and a harder kernel. 

\subsubsection{Lung-RADS}
\label{sec:appendix:data_acquisition-lungrads}
In the screening, reporting and management recommendations were made based on the Lung-RADS classification system~\cite{lungsrads, chelala2021lung, christensen2024acr}.
Initially, Lung-RADS~v1 was used until around April 2023, after which the updated Lung-RADS~v2022 was adopted. 

\subsection{Data curation}
\label{sec:appendix:data_curation}
To identify correct series in the local PACS, each hospital maintained a centralized file that recorded the local patient ID (PID) and the date of the screening exam for participants. 
These local PIDs were mapped to depersonalized IDs (DIDs), which served as the unique identifier across imaging and non-imaging data.

\subsubsection{Imaging data}
A local Orthanc instance, deployed in each participating hospital, acted as a DICOM node to interface with the hospital PACS. 
Using this setup, a Python script, leveraging the PyOrthanc library~\cite{Couture2025}, triggered the local Orthanc to query the PACS via a DIMSE (DICOM Message Service Element) call to retrieve \LCS{} imaging data. For each query, the StudyInstanceUID of the corresponding LDCT study was extracted. 
Once the list of StudyInstanceUIDs was retrieved, another script~\cite{diocmtransfer}, using Orthanc and PyOrthanc, requested the PACS to transfer the identified studies to a dedicated server within the clinical environment. 

This server contains a Karnak~\cite{osirixfoundationKarnak} instance. Karnak de-identified the data following a dedicated de-identification profile. Key de-identification steps included replacing the original patient ID with a salted hash, substituting the patient's name with the corresponding DID, retaining essential metadata, such as study dates and descriptions (to distinguish between baseline and follow-up imaging exams), and completely removing private DICOM tags.

\subsubsection{Non-imaging data} 
Non-imaging information from eligible participants was transferred from the clinical program RedCap database. 
During this transfer, all participant data was de-identified using the DIDs. 
Selected metadata was extracted from the biobank and stored as a json string in the TextValue field of a Structured Reporting (SR) file.

\subsubsection{Inclusion and exclusion criteria}
Only axial CT series were included in \DS{}. 
Image series with a total slice count of <150, non-axial acquisition plane or maximum intensity projection (MIP) reconstructions were excluded.
Furthermore, studies were excluded if no Lung-RADS score and PLCOm2012noRace score was available. Studies had to include at least two axial series (reconstructed with harder and softer kernel) to be included in \DS{}.

\subsection{Data annotation} 
\subsubsection{Lesion characteristics}
\label{app:sec:lesion_characteristics}
For specifying in which of the 18 lung segments (10 for the right lung and 8 for the left lung) the lesion was located, the RadLex terminology~\cite{langlotz2006radlex} was employed for standardization. 
Depending on the selected lesion type (solid, non-solid, cyst), further contextual classification menus were displayed. 
Solid nodules were further characterized as: (1) nodule with complete, central, popcorn, or concentric ring calcification, (2) fat-containing nodule, (4) juxtapleural nodule with smooth margins; and oval, lentiform, or triangular shape, (5)  airway nodule (subsegmental), (6) airway nodule (segmental or more proximal) and (7) spiculation.
No further classification menu was required for non-solid nodules.
Cysts were characterized as: (1) thick-walled cyst (>=2mm - symmetric, asymmetric, or nodular), (2) thin-walled cyst (uniform wall thickness < 2 mm), (3) Unilocular cyst, (4) Multilocular cyst and (5) Cyst with opacity (nodular, ground glass, or consolidation).

\begin{table*}[t]
\centering
\caption{Participants' characteristics at baseline. $^{*}$Data are medians and interquartile ranges (IQRs). 
$^{*1}$ For participants who have stopped smoking.}
\begin{tabular}{l | l  }
\toprule
\rowcolor{Gray}
\textbf{Demographic data} & \\
\hline
Sex  &\\
\rowcolor{Gray}
\multicolumn{1}{r|}{Male} &  878 \\
\multicolumn{1}{r|}{Female} &  800 \\
\hline
\rowcolor{Gray}
Age at \LDCT{}$^{*}$ & 65 (61, 69) \\
\hline
Race or ethnicity &  \\
\rowcolor{Gray}
\multicolumn{1}{r|}{Caucasian} & 1649 \\ 
\multicolumn{1}{r|}{Indigenous} & 18 \\ 
\rowcolor{Gray}
\multicolumn{1}{r|}{Asian} & 4 \\ 
\multicolumn{1}{r|}{Black} & 1 \\ 
\rowcolor{Gray}
\multicolumn{1}{r|}{Hispanic} & 1 \\ 
\multicolumn{1}{r|}{Middle Eastern} & 1 \\ 
\rowcolor{Gray}
\multicolumn{1}{r|}{Other} & 4 \\ 
\hline
\hline
\textbf{Smoking history} & \\
\hline
\rowcolor{Gray}
Smoking status & \\
\multicolumn{1}{r|}{Active} & 1268 \\ 
\rowcolor{Gray}
\multicolumn{1}{r|}{Former} &  410 \\ 
\hline
Number of smoking years$^{*}$ & 45 (40, 50)\\
\hline
\rowcolor{Gray}
Number of years since smoking was stopped$^{*,*1}$ & 7.0 (3.0, 10.75)\\
\hline
\hline
\textbf{Medical history} &\\
\hline
\rowcolor{Gray}
COPD, Emphysema, chronic bronchitis &  \\ 
\multicolumn{1}{r|}{Yes} &  404\\ 
\rowcolor{Gray}
\multicolumn{1}{r|}{No} &   1274\\ 
\hline
PLCOm2012noRace$^{*}$ &  4.2 (2.7, 7.0) \\
\bottomrule
\end{tabular}
\label{tab:description_patient_cohort}
\end{table*}

\begin{table}[]
\centering
\caption{Radiological findings in \DS{}.}
\begin{tabular}{l | l  }
\toprule
 \rowcolor{Gray}
\textbf{LungRADs score} &  \\
\multicolumn{1}{r|}{0} & 51 \\
\rowcolor{Gray}
\multicolumn{1}{r|}{1} & 83 \\
\multicolumn{1}{r|}{2} & 1286 \\
\rowcolor{Gray}
\multicolumn{1}{r|}{3} &  111\\
\multicolumn{1}{r|}{4A} & 94\\
\rowcolor{Gray}
\multicolumn{1}{r|}{4B} &  32\\
\multicolumn{1}{r|}{4X} & 21 \\
\hline
\hline
\rowcolor{Gray}
\textbf{Findings of lung and pleura} &  \\ 
\hline
Emphysema &  \\
\rowcolor{Gray}
\multicolumn{1}{r|}{yes} & 1423 \\ 
\multicolumn{1}{r|}{no} &  127\\ 
\rowcolor{Gray}
\multicolumn{1}{r|}{unknown} &  128\\ 
\hline
\hline
\textbf{Findings in the heart} &  \\ 
\rowcolor{Gray}
\multicolumn{1}{r|}{yes} & 1490  \\ 
\multicolumn{1}{r|}{no} & 110 \\ 
\rowcolor{Gray}
\multicolumn{1}{r|}{unknown} & 78  \\
\hline
Coronary calcification &  \\
\rowcolor{Gray}
\multicolumn{1}{r|}{absent} & 81  \\ 
\multicolumn{1}{r|}{low} & 793 \\ 
\rowcolor{Gray}
\multicolumn{1}{r|}{moderate} & 376  \\
\multicolumn{1}{r|}{severe} & 214  \\
\rowcolor{Gray}
\multicolumn{1}{r|}{unknown} & 214  \\

\bottomrule
\end{tabular}
\label{tab:description_radiological_findings}
\end{table}

\begin{table*}[t]
\centering
\caption{Intra-reader and inter-reader variability in nodule
detection and segmentation. Reader 1 (first reading) served as the references
for the intra-reader and inter-reader detection metrics,
respectively. Dice scores were calculated only for true positive 
nodules. TP = true positive, FN = false negative, FP = false positive. }
\label{tab:reader_agreement}
\begin{tabular}{lcc}
\toprule
\rowcolor{Gray}
\textbf{Metric} & \textbf{Intra-reader} & \textbf{Inter-reader} \\
\midrule
\multicolumn{3}{l}{\textit{Nodule detection}} \\
\rowcolor{Gray}
TP              & 56  & 32   \\
FN  & 6   & 29   \\
\rowcolor{Gray}
FP & 21   & 7   \\
Sensitivity                        & 0.90 & 0.52 \\
\rowcolor{Gray}
Precision                        & 0.73 & 0.82 \\
Mean FP rate per study               & 0.41 & 0.14 \\
\rowcolor{Gray}
Mean FN rate per study                  & 0.12 & 0.57 \\
F1                           & 0.81 & 0.64 \\
\midrule
\rowcolor{Gray}
\multicolumn{3}{l}{\textit{Segmentation overlap (Dice)}} \\
Median (IQR)
  & 0.81 (0.77, 0.87) & 0.73 (0.64, 0.80) \\
  \rowcolor{Gray}
Mean (std)
  & 0.80 (0.09) & 0.72 (0.11) \\
\bottomrule
\label{tab:intra-inter}
\end{tabular}
\par\medskip
\begin{minipage}{\textwidth}
\end{minipage}
\end{table*}

\subsubsection{Reader variability}
To access intra- and inter-reader variability, a subset of 51 baseline studies with varying Lung-RADS scores was again annotated by the annotating radiologist (reader 1) with a wash-out period of at least 4 months (reader 1*) and additionally by a first-year resident in radiology (reader 2). \\
The inter- and intra-reader analysis was performed for pulmonary nodules and not for cysts to allow better comparison with variability values reported in previous studies~\cite{armato2009assessment, yip2017application}. 
Series with only nodules $< 34\,\mathrm{mm}^3$ or no findings in the reference (reader 1) and the comparison reading (reader 1*/reader 2) were regarded as a true negative (TN) series. 
An overlap-based matching criterion between nodules of reader 1 and reader 1*/reader 2 was utilized to assess the nodule detection. Nodules were regarded as matched with a $Dice=\frac{2\times\text{Volume of intersection}}{\text{Total volume}}> 0$. Multiple nodules were allowed to match the same reference lesion. In such cases, the matching nodule with the largest Dice was selected and the others were disregarded (neither treated as true positives (TPs) or false positives (FPs)).
A matched nodule was counted as a TP if the nodule volume was  $\geq 34\,\mathrm{mm}^3$ in either read. 
Matched nodules were disregarded and not counted as TPs if the nodule had a volume smaller than $34\,\mathrm{mm}^3$ in both reads. 
Reference nodules smaller than $34\,\mathrm{mm}^3$ with no match in the second read were not counted as false negatives (FNs). 
Likewise, unmatched comparison-read nodules smaller than $34\,\mathrm{mm}^3$ were not counted as FPs.

The number of TP, FP and FN nodules were reported. 
Additionally, the sensitivity ($\frac{TP}{TP + FN}$), precision ($\frac{TP}{TP + FP}$), mean false-positive rate per study ($\frac{\sum_{\text{studies}}\text{number of FPs per study}}{\text{number of studies}}$), mean false-negative rate per study ($\frac{\sum_{\text{studies}}\text{number of FNs per study}}{\text{number of studies}}$), as well as the nodule-level F1 score ($\frac{2TP}{2TP + FP + FN}$) were calculated. 
Sensitivity and precision, as well as the mean FP and FN rates per scan, describe equivalent performance characteristics, differing only in the reference standard used: Reader 1 or Reader 1$^{*}$/Reader 2. 
For TP nodules, the intra/inter-reader segmentation performance was evaluated by calculating the mean (standard deviation (std)) as well as the median (0.25-quantile, 0.75-quantile) Dice score. 
Furthermore, for the TP cases, the intra- and inter-reader nodule classification performance was assessed by comparing the number of alike and different answers for the lesion type, including solid, ground-glass, part-solid nodules.

\subsection{Data analysis}
The lesion volume analysis with PyRadiomics~\cite{van2017computational} serves as a basic example of how data analysis code can be applied to DICOM objects within PARADIM’s managed execution environment.

The platform enables systematic processing of medical imaging data in a user-friendly model-to-data approach with minimal setup effort. Users simply package their code into a container that complies with the Open Container Initiative (OCI) specifications and push it to a docker registry. 
The container must be executable (at the series or study level), and all outputs must be stored as DICOM objects with the container's name as the series description for traceability. 
In PARADIM, the containerized application is converted into an Apptainer image. 
The DICOM outputs are automatically assigned to the corresponding Kheops collection. 
Results (stored as a flat dictionary in the TextValue field of an SR file) can be aggregated across the collection, exported as a single CSV file, and provided to the user for further analysis.

\raggedright
\section{Results}
\subsection{Reader variability}
\begin{figure*}[t!]
\subfloat[]{%
    \includegraphics[width=.48\linewidth]{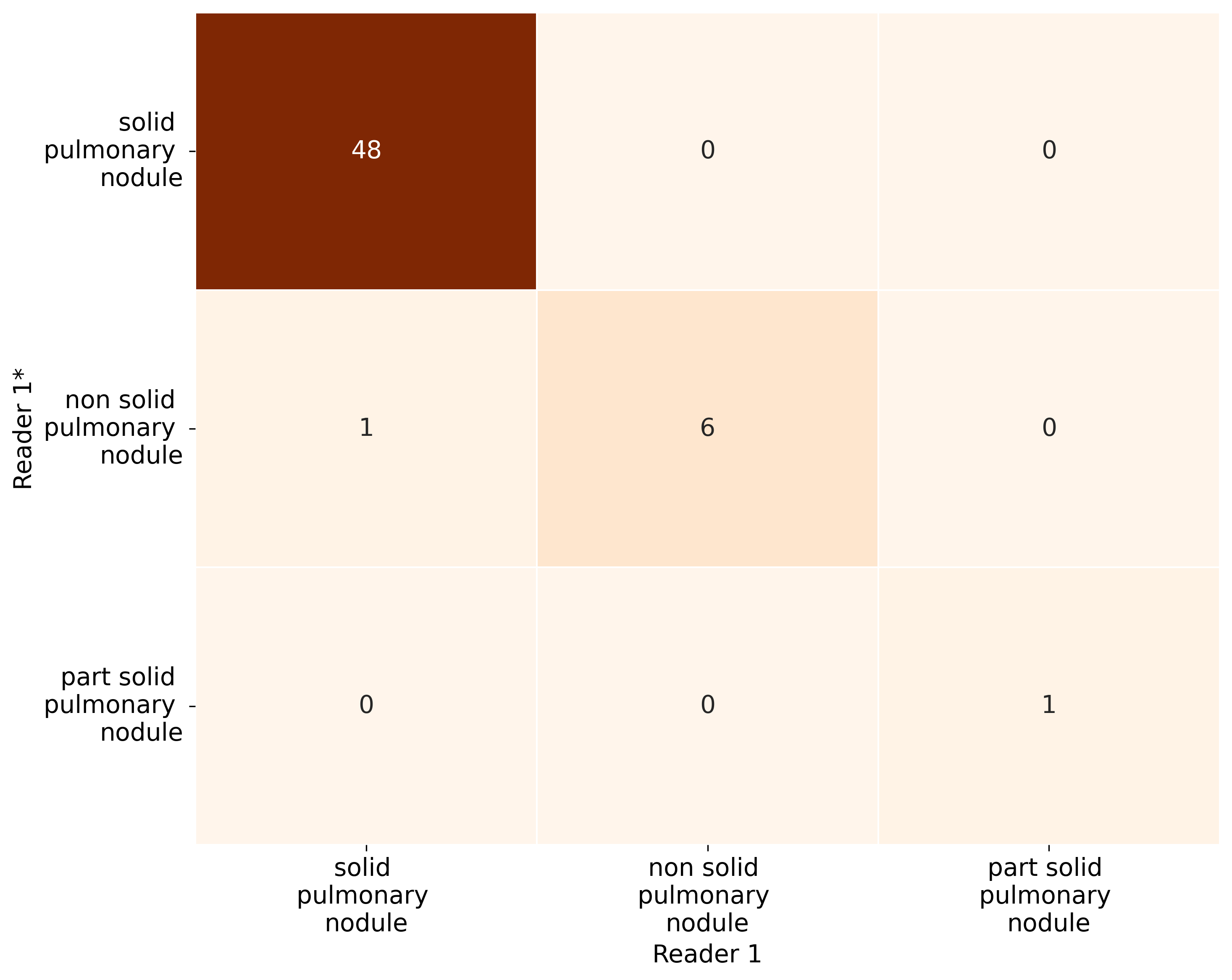}%
    \label{subfig:app:interreader_findings_intrareader_confusion_matrix}%
}\hfill
\subfloat[]{%
    \includegraphics[width=.48\linewidth]{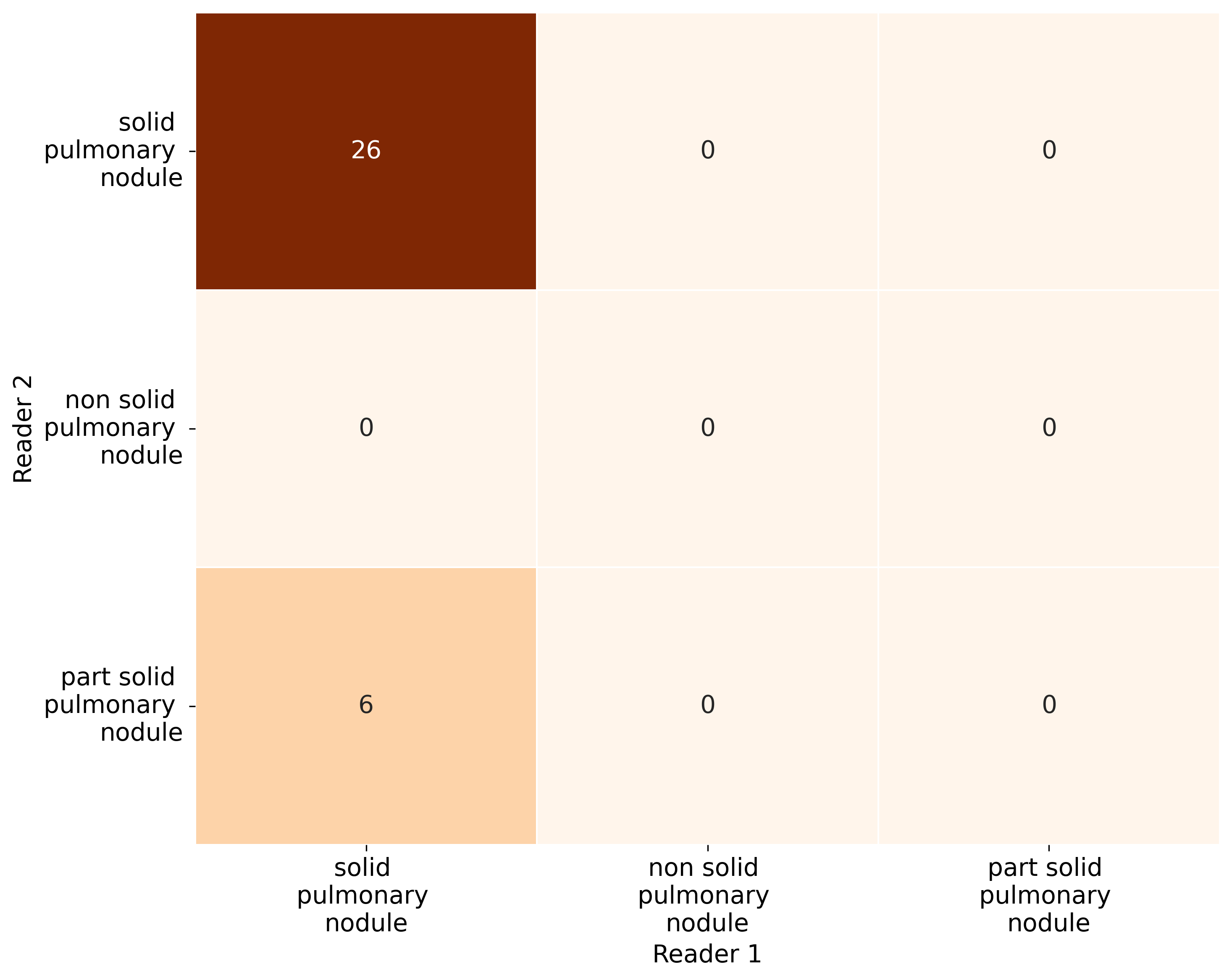}%
    \label{subfig:app:interreader_findings_interreader_confusion_matrix}%
}\hfill \\
\subfloat[]{%
    \includegraphics[width=.48\linewidth]{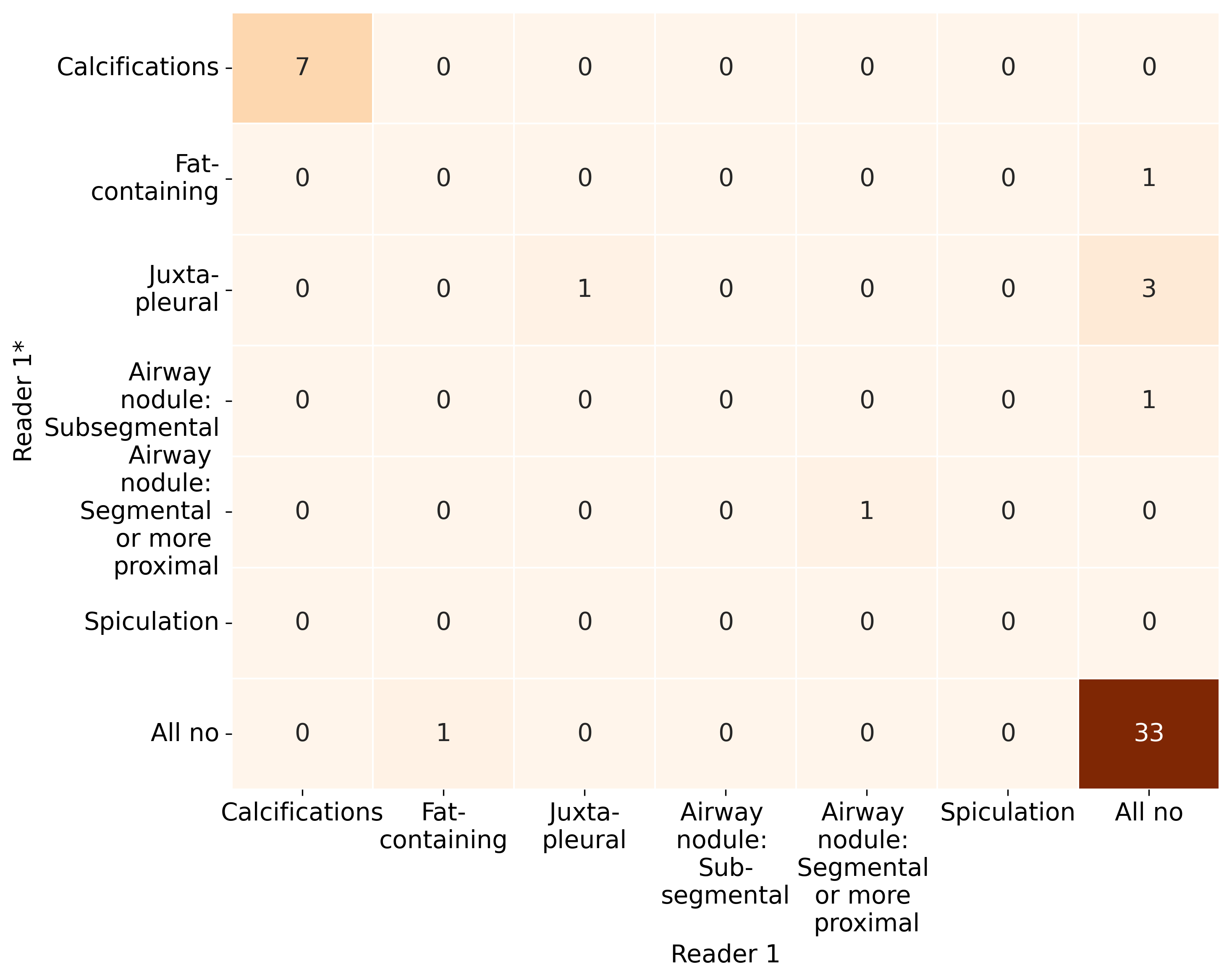}%
    \label{subfig:app:subfig:app:intrareader_findings_intrareader_confusion_matrix_characteristics}%
}\hfill
\subfloat[]{%
    \includegraphics[width=.48\linewidth]{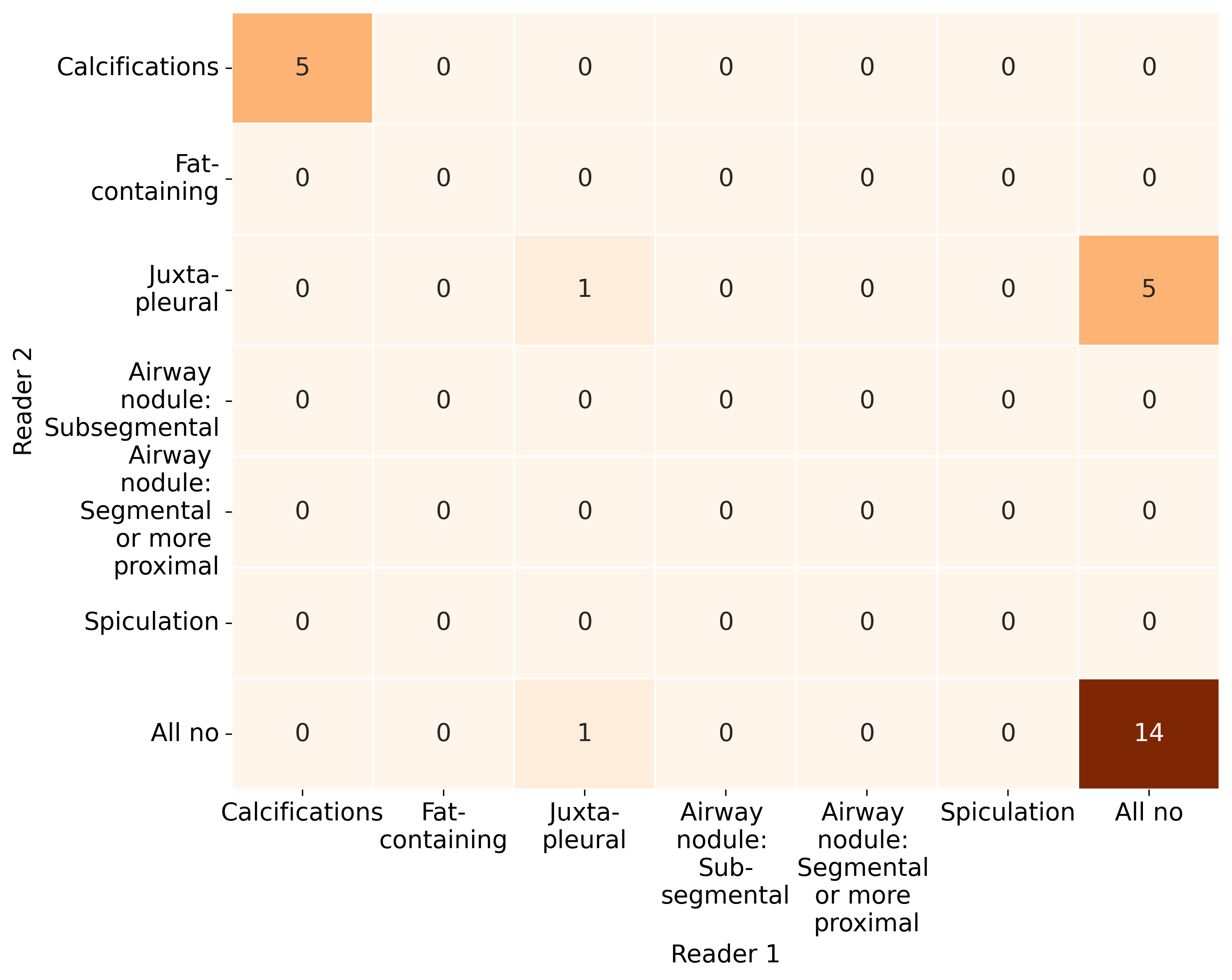}%
    \label{subfig:app:subfig:app:interreader_findings_intrareader_confusion_matrix_characteristics}%
}\hfill
\caption{(a,c) Intra-reader analysis; (b,d) inter-reader analysis. (a)/(b) Number of reported nodule characteristics for the pulmonary nodules detected in both readings. (c)/(d) Comparison of subcharacteristics for the nodules that were in both readings classified as solid.}
\label{tab:app:res:reader_char}
\end{figure*}

Intra- and inter-reader analysis results are summarized in Table~\ref{tab:intra-inter}. 
The intra-reader analysis showed that 56 nodules were detected in both readings, 6 nodules were annotated only in the first reading and 21 nodules were only detected in the second reading. 
Sensitivity, precision, mean FP rate per study, mean FN rate per study, and F1 score in the intra-reader analysis was 0.90, 0.73, 0.41, 0.12, and 0.81 respectively. 
For the cases that were annotated in both reading, a median (IQR)  and mean (std) Dice score of 0.81 (0.77, 0.87) and 0.8 (0.09) was determined, respectively. 
Most of these nodules were classified as solid in both readings ($n=48$) (Fig.~\ref{tab:app:res:reader_char}(a)). \\
In the inter-reader analysis 32 nodules were annotated by both readers, while 29 nodules were only annotated by reader 1 and 7 only by reader 2. The inter-reader sensitivity, precision, mean FP rate per study, mean FN rate per study and F1 score was 0.52, 0.82, 0.14, 0.57, and 0.64, respectively. 
For the nodules, annotated by both readers a median (IQR) and mean (std) Dice  of 0.73 (0.64, 0.80) and 0.72 (0.11) was calculated, respectively. 
Nodule characterizations coincided in 81\% (Fig.~\ref{tab:app:res:reader_char} (b)). \\

\begin{figure*}[t!]
\subfloat[]{%
    \includegraphics[width=.49\linewidth]{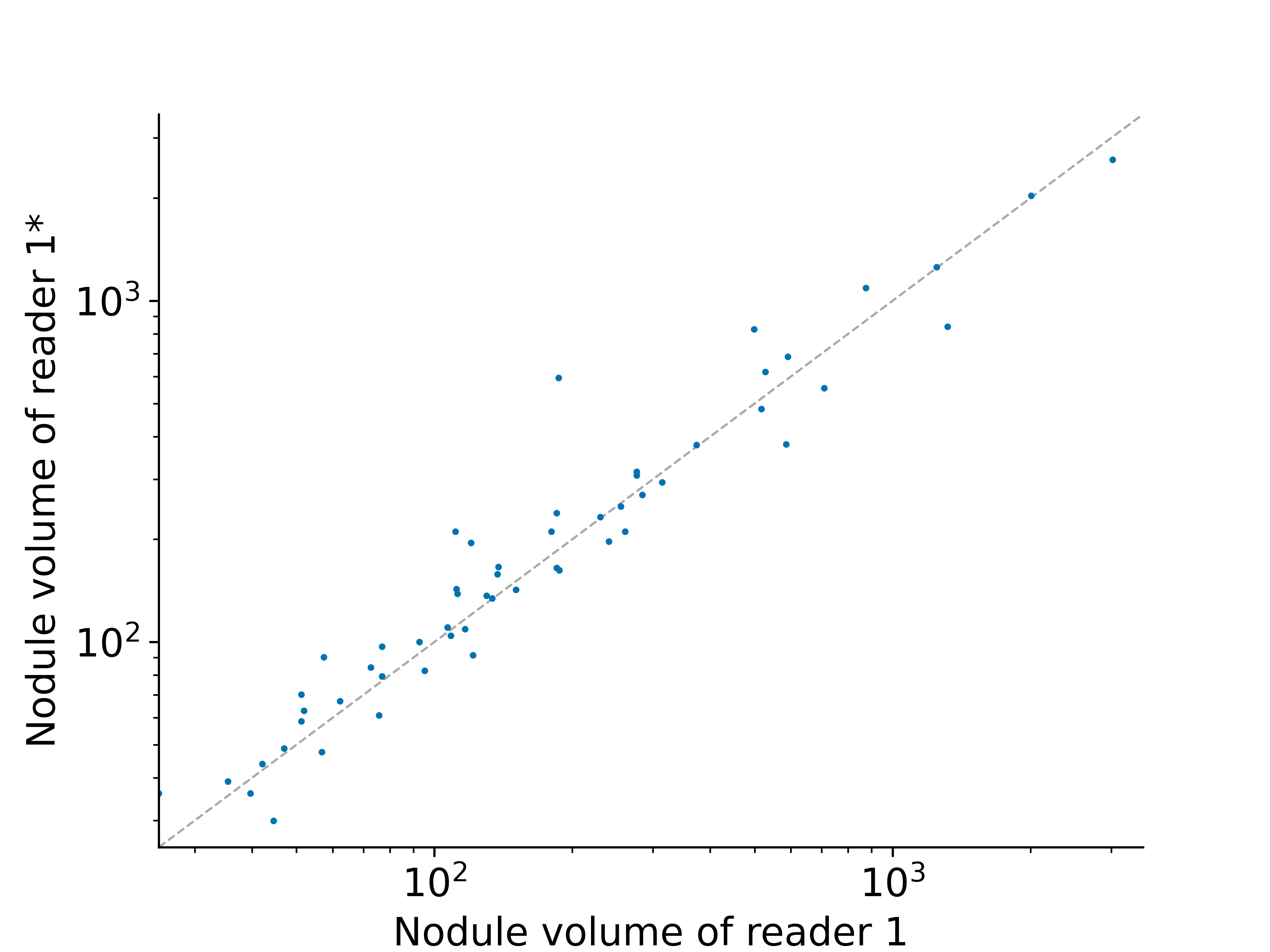}%
    \label{subfig:app:volume_reader_a}%
}\hfill
\subfloat[]{%
    \includegraphics[width=.49\linewidth]{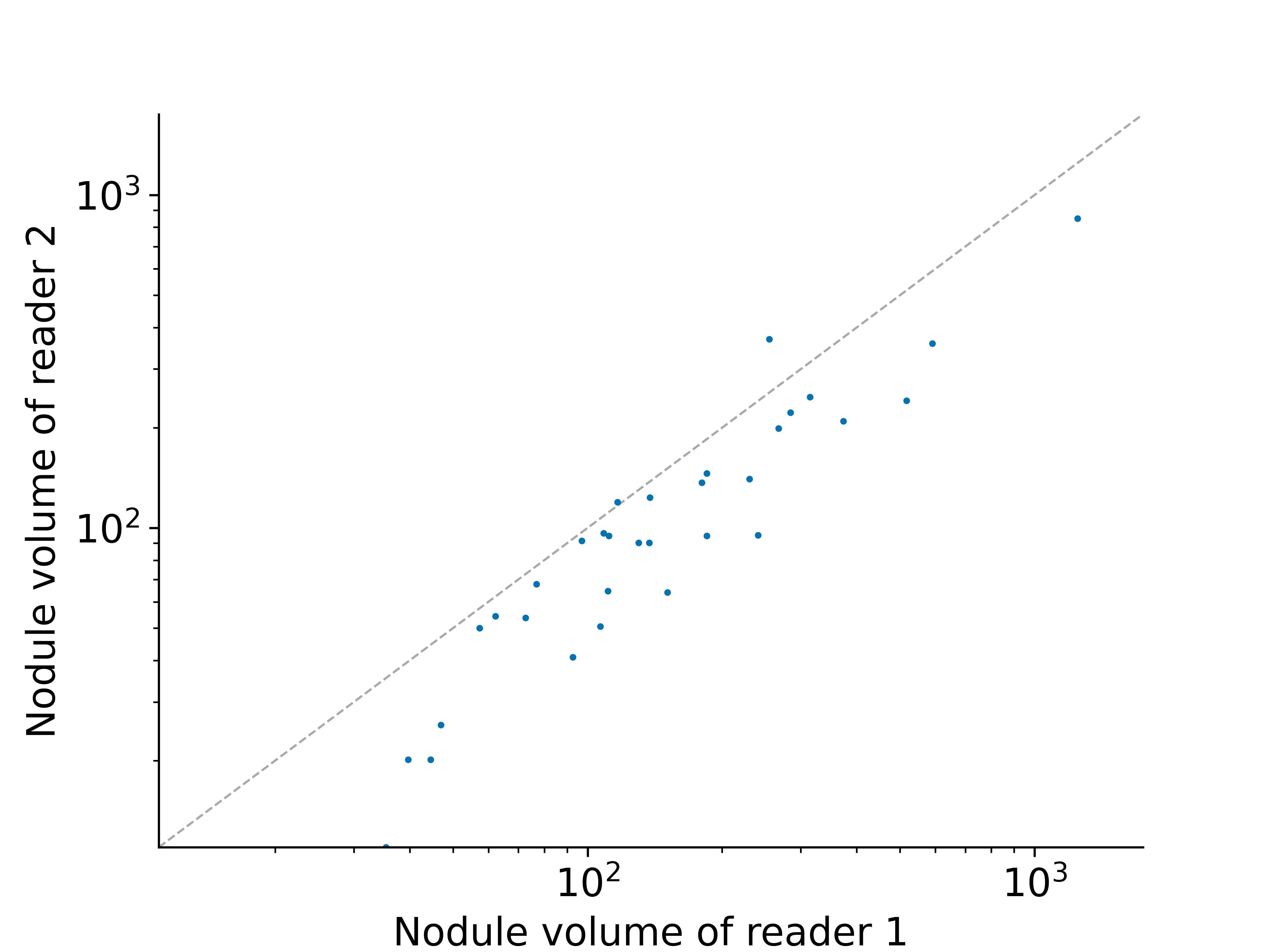}%
    \label{subfig:app:volume_reader_b}%
}\hfill
\caption{Volume of findings in $\mathrm{mm}^3$ segmented in both readings of (a) intra-reader analysis  and (b) inter-reader analysis.}
\label{fig:app:volume_reader}
\end{figure*}

\section{Discussion}
\subsection{Other accessible lung nodule CT dataset}
Datasets~\cite{shaukat2019computer}  such as ANODE09~\cite{van2010comparing}, LungCT-diagnosis~\cite{grove2015quantitative}, SPIE-AAPM~\cite{armato2015lungx}, RIDER lung CT~\cite{zhao2009evaluating}, RIDER PET-CT~\cite{muzi2015data}, VIA/ELCAP~\cite{henschke2001early}, Public lung database to address drug response~\cite{reeves2009public}, QIN lung CT~\cite{kalpathy2015qin, goldgof2015data}, Lung CT segmentation challenge 2017~\cite{yang9data}, LNDb~\cite{pedrosa2019lndb} 
are not listed in Table 1 due to their low number of studies (< 300). 
UniToChest~\cite{chaudhry2022unitochest} is not included due to missing \LCS{} context.
Datasets such as DeepLesion~\cite{yan2018deeplesion} that did not provide full CT volumes were also not included for comparison in Table 1. 

\subsection{Reader variability}
Armato et al.~\cite{armato2009assessment} reported a mean sensitivity between 0.51-0.83 and a mean FP rate between 0.33-1.39 per case when comparing the radiologist detections to different panel answers.  
Our inter-reader sensitivity/precision of 0.52/0.82 and mean FN rate of 0.57 per study were in line with these results, while our FP rate per study of 0.14 was slightly better. 
The median (std) inter-reader Dice score of  0.73 (0.64, 0.80), respectively, were slightly lower with overlapping IQRs with a previously reported median (IQR) inter-reader Dice coefficient of  0.82 (0.77, 0.85)~\cite{yip2017application}. 
This highlights that variation in annotations is to be expected and should be considered when evaluating AI models.

\clearpage
\raggedbottom
\printbibliography
\pagebreak
\clearpage

\includepdf[
  pages=1,
  pagecommand={
    \begin{tikzpicture}[remember picture,overlay]
      \node[anchor=north west]
        at ([xshift=0.93in,yshift=-0.71in]current page.north west)
        {\Large\bfseries S4 Annotation guide};
    \end{tikzpicture}
  }
]{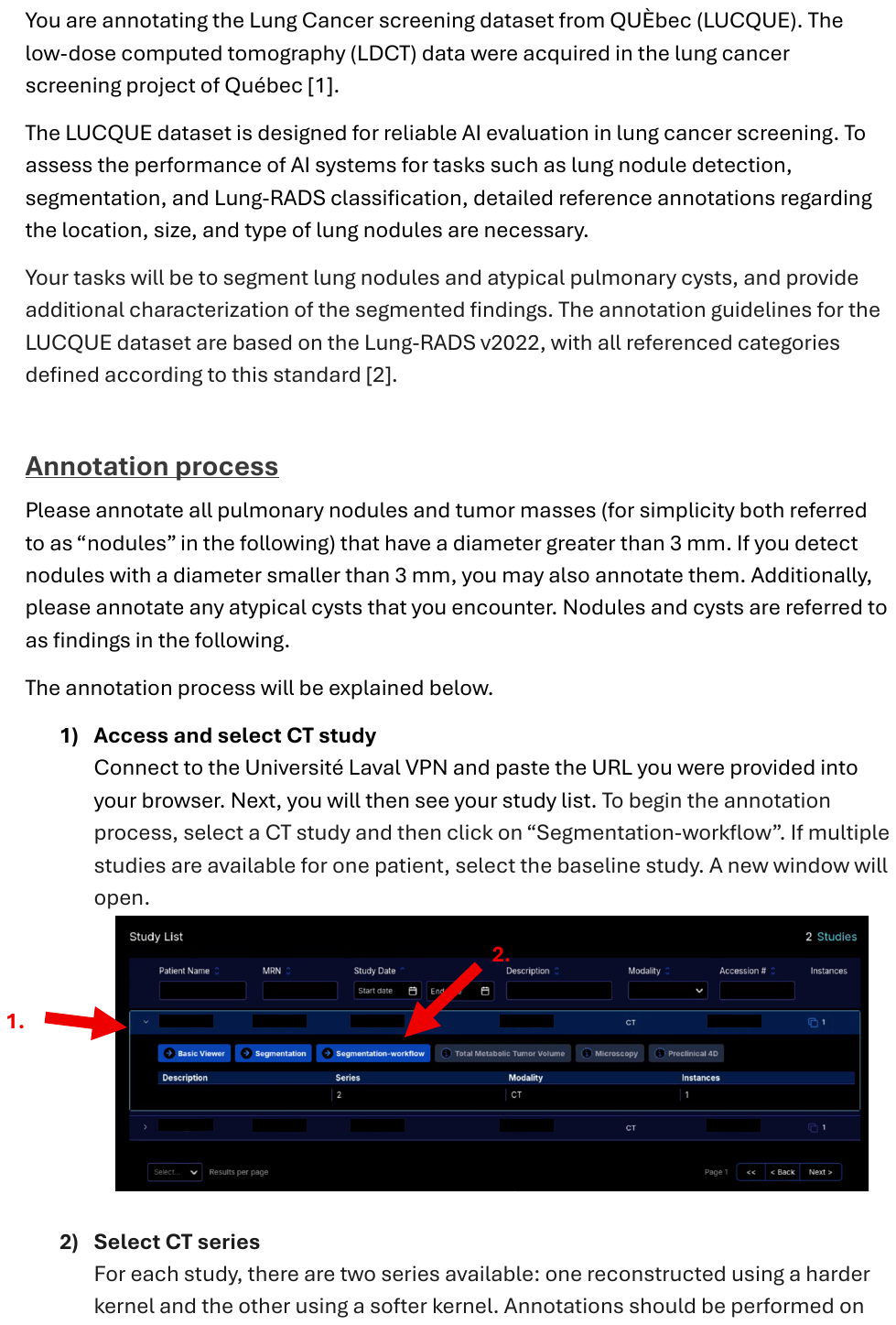}
\includepdf[pages=2-6]{annotation_guidelines_lucque.pdf}
\end{refsection}
\end{document}